\documentclass{egpubl}
\usepackage{pg2025}
\usepackage{xcolor}
\usepackage{needspace}

\SpecialIssuePaper         

\CGFStandardLicense

\usepackage[T1]{fontenc}
\usepackage{dfadobe}  

\usepackage{booktabs} 
\usepackage[linesnumbered,ruled,vlined]{algorithm2e}
\usepackage{algorithmicx}
\usepackage{algpseudocode}
\usepackage{setspace}
\usepackage{wrapfig}
\usepackage{pifont}
\newcommand{\cmark}{\ding{51}}%
\newcommand{\xmark}{\ding{55}}%

\SetKwInput{KwInput}{Input}
\SetKwInput{KwOutput}{Output}

\usepackage[ruled]{algorithm2e} 

\SetAlFnt{\small}
\SetAlCapFnt{\small}
\SetAlCapNameFnt{\small}
\SetAlCapHSkip{0pt}

\usepackage{physics}
\usepackage{float}
\usepackage{stfloats}  
\usepackage{multirow}

\SetCommentSty{mycommfont}

\usepackage{cite}  
\BibtexOrBiblatex
\electronicVersion
\PrintedOrElectronic
\ifpdf \usepackage[pdftex]{graphicx} \pdfcompresslevel=9
\else \usepackage[dvips]{graphicx} \fi

\usepackage{egweblnk} 

\usepackage{amssymb}
\usepackage{multibib}
\newcites{supp}{References (Supplementary)}
\title[Parallel Mesh Optimization for Intersection-Free Low-Poly Modeling on the GPU]%
      {PaMO: Parallel Mesh Optimization for Intersection-Free Low-Poly Modeling on the GPU}

\author[S. Oh, X. Yuan, X. Wei, R. Shi, F. Xiang, M. Liu \& H. Su]
{\parbox{\textwidth}{\centering
Seonghun Oh$^{1}$\thanks{Equal contribution}\orcid{0009-0003-1578-9446},
Xiaodi Yuan$^{2}$\footnotemark[1]\orcid{0009-0003-2320-2297},
Xinyue Wei$^{2}$\footnotemark[1]\orcid{0000-0002-9466-6836},
Ruoxi Shi$^{2}$\orcid{0009-0002-2502-1329},
Fanbo Xiang$^{3}$\orcid{0009-0005-5335-873X},
Minghua Liu$^{3}$\orcid{0000-0002-6413-023X},
Hao Su$^{2,3}$\thanks{Corresponding author}\orcid{0000-0002-1796-2682}}
\\
{\parbox{\textwidth}{\centering
$^{1}$Yonsei University\quad
$^{2}$University of California, San Diego\quad
$^{3}$Hillbot Inc.
}
}
}

\begin{document}

\teaser{
  \includegraphics[width=\textwidth]{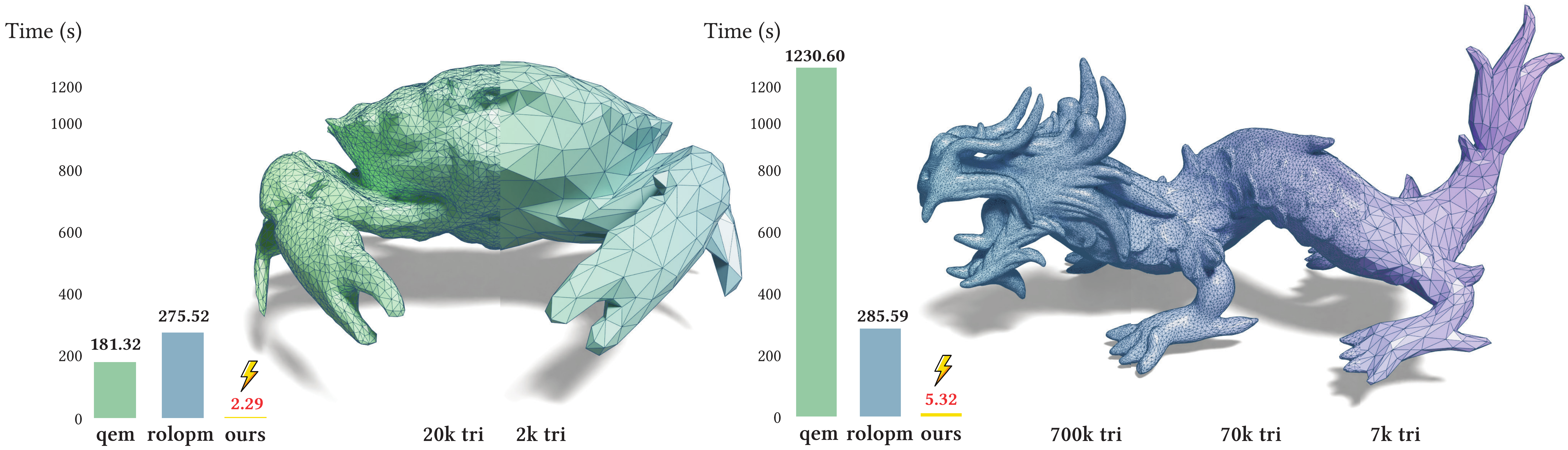}
  \centering
  \vspace{-2em}
  \caption{We propose a novel GPU-based mesh optimization method to convert meshes in the wild into low-poly, intersection-free, manifold meshes. Our algorithm efficiently processes large-scale meshes in seconds, preserving self-intersection-free and manifold properties with high quality. \textbf{Left:} Reducing the 2M-face ``crab'' to 0.1\% in 2.29s. \textbf{Right:} Reducing the 7M-face ``dragon'' to 0.1\% in 5.32s. (Only the output meshes are shown above).}
  \label{fig:teaser}
}

\maketitle
\begin{abstract}
   Reducing the triangle count in complex 3D models is a basic geometry preprocessing step in graphics pipelines such as efficient rendering and interactive editing. However, most existing mesh simplification methods exhibit a few issues. Firstly, they often lead to self-intersections during decimation, a major issue for applications such as 3D printing and soft-body simulation. Second, to perform simplification on a mesh in the wild, one would first need to perform re-meshing, which often suffers from surface shifts and losses of sharp features. Finally, existing re-meshing and simplification methods can take minutes when processing large-scale meshes, limiting their applications in practice. To address the challenges, we introduce a novel GPU-based mesh optimization approach containing three key components: (1) a parallel re-meshing algorithm to turn meshes in the wild into watertight, manifold, and intersection-free ones, and reduce the prevalence of poorly shaped triangles; (2) a robust parallel simplification algorithm with intersection-free guarantees; (3) an optimization-based safe projection algorithm to realign the simplified mesh with the input, eliminating the surface shift introduced by re-meshing and recovering the original sharp features. The algorithm demonstrates remarkable efficiency, simplifying a 2-million-face mesh to 20k triangles in 3 seconds on RTX4090. We evaluated the approach on the Thingi10K dataset and showcased its exceptional performance in geometry preservation and speed. \url{https://seonghunn.github.io/pamo/}


\end{abstract}  

\section{Introduction}

3D meshes play a crucial role across various domains, including gaming, CAD design, AR/VR, and physical simulation. They are commonly obtained through scanning~\cite{calli2015benchmarking,downs2022google}, 3D reconstruction~\cite{wei2023neumanifold,tang2022nerf2mesh}, and 3D generation techniques~\cite{lin2023magic3d,liu2023one}. However, these meshes typically contain a large number of polygons to express detailed geometry, posing a considerable challenge for integration into many applications due to their memory and computational expense. Therefore, finding a low-poly approximation is necessary. This problem, known as \emph{mesh simplification}, is a long-standing topic, though many challenges remain unsolved. 
Mesh simplification should preserve the topological structures (e.g., manifoldness) and overall contour of the original mesh while avoiding defects such as self-intersections. Existing methods have limitations that can be categorized into three areas. \textit{Failure to preserve topological structures:} Vertex clustering methods~\cite{low1997model, rossignac1993multi} can introduce non-manifold structures. \textit{Inability to handle arbitrary inputs:} Edge-collapsing methods~\cite{hoppe1996progressive} do preserve the original topology but require a manifold input mesh. Although non-manifold meshes can be converted into manifolds using remeshing techniques, this often results in surface shifts and the loss of sharp features. For instance, a common remeshing strategy~\cite{museth2013openvdb} converts input triangle soups into unsigned distance fields (UDF) and then extracts surfaces with an extra margin, resulting in a slightly expanded surface compared to the original mesh. \textit{Generation of self-intersections:} Most mesh simplification methods may introduce self-intersections during the simplification process, greatly limiting their use in many downstream applications that cannot tolerate self-intersections, such as boolean operations and soft-body simulations. While~\cite{chen2023robust} addresses the self-intersection issue, it significantly slows down the process, taking tens of minutes to process a single mesh.

Many downstream applications, such as interactive design, demand high efficiency from mesh processing. However, it can take existing algorithms a considerable amount of time to process a large 3D model, which is not acceptable. While the immense parallel computation power of modern GPUs is promising, the majority of existing algorithms are designed for sequential execution and require complex modifications for GPU adaptation. For instance, OpenVDB~\cite{museth2013openvdb} supports a CPU-based parallel remeshing algorithm, but its GPU-based parallel counterparts remain unexplored. Some efforts~\cite{gautron2023interactive} focus on parallelizing edge collapse; however, they do not succeed in preventing self-intersections.




In this paper, we propose a novel GPU-based mesh optimization algorithm to address the three main limitations of existing mesh simplification methods. This algorithm not only reduces the number of triangles in the input mesh but also optimizes arbitrary mesh inputs into manifold, intersection-free results with well-shaped triangles.
The pipeline consists of three stages: remeshing, simplification, and safe projection. These stages are meticulously designed to complement each other. The remeshing stage takes arbitrary meshes as input and converts them into manifold meshes with well-shaped triangles, which also enhances the efficiency of the simplification algorithm (see Figure~\ref{fig:undo_remesh}). The simplification stage iteratively collapses edges and reverts collapse operations that cause intersections, as detected by our proposed fast and robust GPU-based triangle intersection test. We have specially designed algorithms to handle intersections between vertex-sharing triangles, which most existing intersection detection methods rarely consider but are very common in our use case. 
The safe projection stage optimizes the vertex positions to compensate for the surface shift introduced by remeshing and simplification (see Figure~\ref{fig:sharp_feat}). We formulate the safe projection as a constrained optimization problem, where the objective is to minimize mesh distances to the input and mesh quality losses, subject to an intersection-free constraint. Notably, this optimization problem resembles hyperelastic thin shell simulations in finite-element method (FEM) frameworks. Drawing inspiration from the intersection-free FEM solver~\cite{Li2020IPC}, we have designed a GPU-based Newton-type optimizer to effectively solve the safe projection problem.


Our key contributions can be summarized as follows: \textbf{(a) Fully GPU-based mesh optimization: } To the best of our knowledge, we have designed the first fully GPU-based mesh optimization algorithm capable of converting arbitrary input meshes into intersection-free, low-poly manifold meshes with high efficiency and quality. Our method achieves exceptional performance, simplifying a 2-million-face mesh to 2k triangles with faithful detail in just 2.75 seconds on an RTX 4090 GPU. \textbf{(b) Robust intersection-free simplification algorithm:} We introduce a parallel, GPU-based simplification algorithm that guarantees self-intersection-free results. We specifically handle the detection of intersections in vertex-sharing triangles, which most existing detection algorithms rarely address. \textbf{(c) Physically inspired safe projection: } We propose a physically inspired safe projection algorithm to mitigate surface shifts and the loss of sharp features while avoiding intersections.

We evaluate our method using the Thingi10K dataset~\cite{Thingi10K} and compare it with state-of-the-art baseline methods. Our results demonstrate that our method consistently produces low-poly meshes that closely resemble the input meshes and possess superior geometric properties. Additionally, it is significantly faster than most existing baselines, decimating 98\% of the meshes in Thingi10K to 1\% of their original number of triangles within 2 seconds. 
These advantages make our proposed method an attractive option for integration into graphics pipelines and various downstream applications that require high-quality, low-poly meshes in time-sensitive environments. See Figure~\ref{fig:teaser} for some examples.

\begin{figure*}[t]
\begin{center}
   \includegraphics[width=\linewidth]{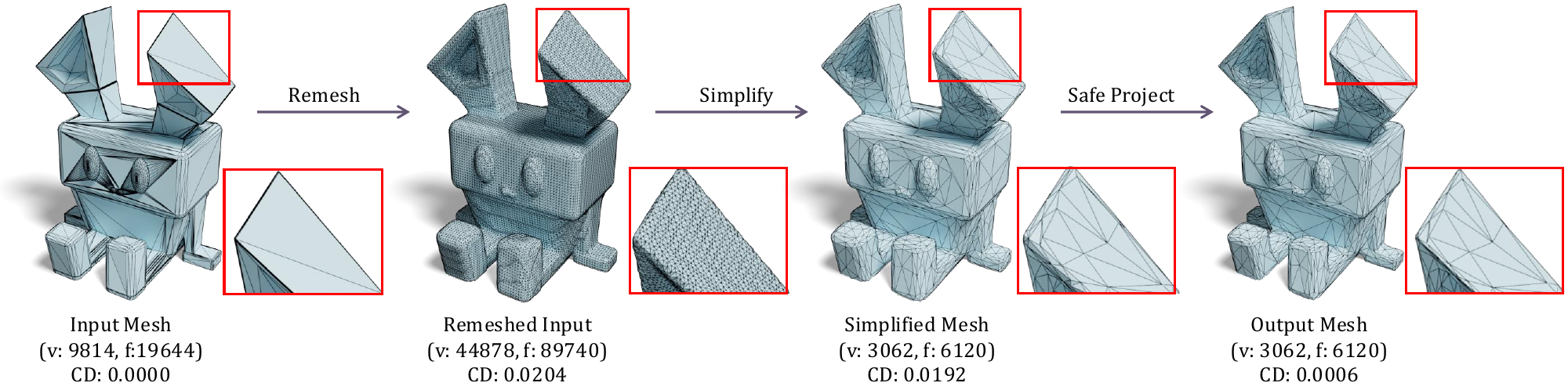}
\end{center}
  \caption{\textbf{Overall pipeline of our method.} Taking an arbitrary mesh as input, our GPU-based method first remeshes it into a manifold, intersection-free mesh. It is then simplified into a low-poly mesh with the desired face count. Finally, we project the surface vertices back towards the original mesh to mitigate surface shifts and recover sharp features. CD denotes the chamfer distance between the input mesh and the generated mesh of each stage.}
    \vspace{-1em}
\label{fig:pipeline}
\end{figure*}


\section{Related Work}
\subsection{Remeshing}

Remeshing algorithms transform an arbitrary input mesh into a new mesh with desired properties, such as being manifold or intersection-free, while also preserving the original mesh's features like surface contours and sharp edges. Most remeshing methods can be divided into two categories: \emph{mesh repairing} and \emph{volumetric remeshing}. 


\emph{Mesh repairing} algorithms retain input vertices. Early methods for mesh repairing addressed specific mesh defects using local operations, such as hole filling~\cite{podolak2005atomic}, gap closing~\cite{bischoff2005structure}, and intersection removal~\cite{attene2010lightweight,attene2014direct}. However, repairing one type of flaw often introduces new flaws of different types. More recently, robust methods have been proposed that extract tetrahedral meshes from triangle soup inputs~\cite{hu2018tetrahedral,hu2020fast}, although these meshes may not always be manifold.

\emph{Volumetric remeshing} generates new meshes without reusing original vertices by first converting the input mesh to a volumetric representation, typically a signed or unsigned distance field, and then extracting the isosurface from that distance field.


Several strategies have been proposed to reduce the computational workload when calculating unsigned distance fields from a triangle mesh. These include the use of BVH trees, as used in Open3D~\cite{zhou2018open3d}; BFS-based voxelization, as seen in OpenVDB~\cite{museth2013openvdb}; Embree~\cite{wald2014embree} and KD-Tree-based filtering, as implemented in Trimesh~\cite{trimesh}. However, these methods are not particularly efficient in environments with extensive GPU parallelism. Our work introduces a voxel grid hierarchy method that demonstrates remarkable efficiency on GPUs. 

Isosurface extraction algorithms have evolved significantly since the introduction of the marching cubes method~\cite{lorensen1987marching}. Various advancements have improved the quality of extracted surfaces~\cite{natarajan1994generating,tcherniaev1996marching,doi1991efficient,nielson2004dual}, typically ensuring that they are manifold and intersection-free, but often at the expense of preserving sharp features. While extended marching cubes~\cite{kobbelt2001feature}, dual contouring~\cite{ju2002dual}, and dual-grid marching cubes~\cite{schaefer2005dual} preserve sharp features, they can potentially generate intersecting faces. More recent approaches, including data-driven isosurface extraction methods~\cite{liao2018deep,chen2022neural,chen2021neural}, utilize neural networks to enhance surface prediction. Additionally, differentiable isosurface extraction techniques such as DMTet~\cite{shen2021deep}, FlexiCubes~\cite{shen2023flexible}, and DiffMC~\cite{wei2023neumanifold} integrate surface extraction directly into optimization workflows. Our contribution extends this body of work by providing a highly efficient GPU implementation of the dual marching cubes algorithm.

\subsection{Mesh Simplification}

Mesh simplification reduces an input mesh to fewer vertices and faces. One of the earliest approaches, vertex clustering~\cite{rossignac1993multi,low1997model}, merges mesh vertices based on spatial proximity and then removes degenerate faces resulting from the merge. Although this method allows for massive parallelization, it does not preserve the mesh topology or the manifold property. Edge collapsing~\cite{hoppe1996progressive} addresses this by incrementally merging edges while ensuring that each operation maintains the manifold property by incorporating conditions such as the link condition ~\cite{dey1999topology}. The edge collapsing paradigm has inspired extensive research on the criteria for edge removal~\cite{garland1997surface,lindstrom2000image,wei2010feature,borouchaki2005simplification,lescoat2020spectral,gumhold2003intersectiona,chen2023robust}. However, edge collapsing presents significant challenges for parallelization. 

Extensive research has been conducted on parallel simplification to address the aforementioned challenges. Notably, methods based on vertex removal~\cite{odaker2016gpu} and vertex clustering techniques~\cite{decoro2007real,legrand2015morton} achieve very high speeds. However, these approaches often compromise mesh quality, struggling to preserve sharp features and manifold properties.
Some works utilize the parallelism of GPUs to perform iterative edge collapses using Quadric Error Metrics (QEM). They focus on identifying independent regions to prevent operational collisions during each parallel local collapsing operation~\cite{papageorgiou2015triangular, mousa2021high, koh2018gpu, lee2016parallel}.
While some of these studies use link conditions to preserve the manifold properties of decimated meshes, they are prone to race conditions in defining such regions. These studies sometimes struggle to produce high-quality meshes without numerous iterations.  


Recent research by Jiang et al.~\cite{jiang2022declarative} introduces a parallel algorithm that employs priority queues and mutex-based synchronization to prevent race conditions and ensure consistency. However, this method is confined to CPU multithreading and fails to utilize the vast computational power of GPUs. Subsequent research by Gautron et al.~\cite{gautron2023interactive} leverages lock-free parallel cost propagation to identify independent regions and prevent race conditions, significantly reducing the algorithm's runtime. 
Nonetheless, this method does not ensure certain mesh properties, particularly the prevention of self-intersections. To address this, we build on the work of Gautron et al.~\cite{gautron2023interactive} and propose a novel technique that uses fully parallelized GPU self-intersection detection to revert any operations that cause intersections, enhancing the quality of the resulting mesh.

\subsection{Triangle Intersection Detection}

Möller's algorithm \cite{moller1997fast} is widely used for ray-triangle intersection checks and has also been extended to test triangle-triangle intersections. Following the same general scheme, Devillers' algorithm \cite{devillers2002faster} builds on this by eliminating intermediate explicit constructions and relying solely on the signs of $4\times 4$ determinants, which enhances numerical robustness and efficiency. Particularly, for coplanar triangles, Devillers' algorithm begins by checking if a vertex of one triangle is inside the other, indicating an intersection; if not, the algorithm proceeds to evaluate potential line-segment intersections between the triangles. Despite its robustness and efficiency, Devillers' algorithm assumes that the two triangles are independent, i.e., they do not share common vertices. Under this assumption, any overlapping vertices would result in a detected intersection. However, this assumption does not hold in the scenario of mesh self-intersection detection, where it is crucial to differentiate between true self-intersections and legitimate cases of vertex- or edge-sharing (see Figure \ref{fig:insect_cases_2D}). To address this issue, we introduce a case-based triangle intersection test specifically designed for detecting self-intersections in meshes. Building upon Devillers' algorithm, we specifically address coplanar vertex-sharing triangle pairs and edge-sharing triangle pairs according to the number of shared vertices, thereby improving the robustness of self-intersection detection.
\begin{figure}[t]
    \centering
    \includegraphics[width=\linewidth]{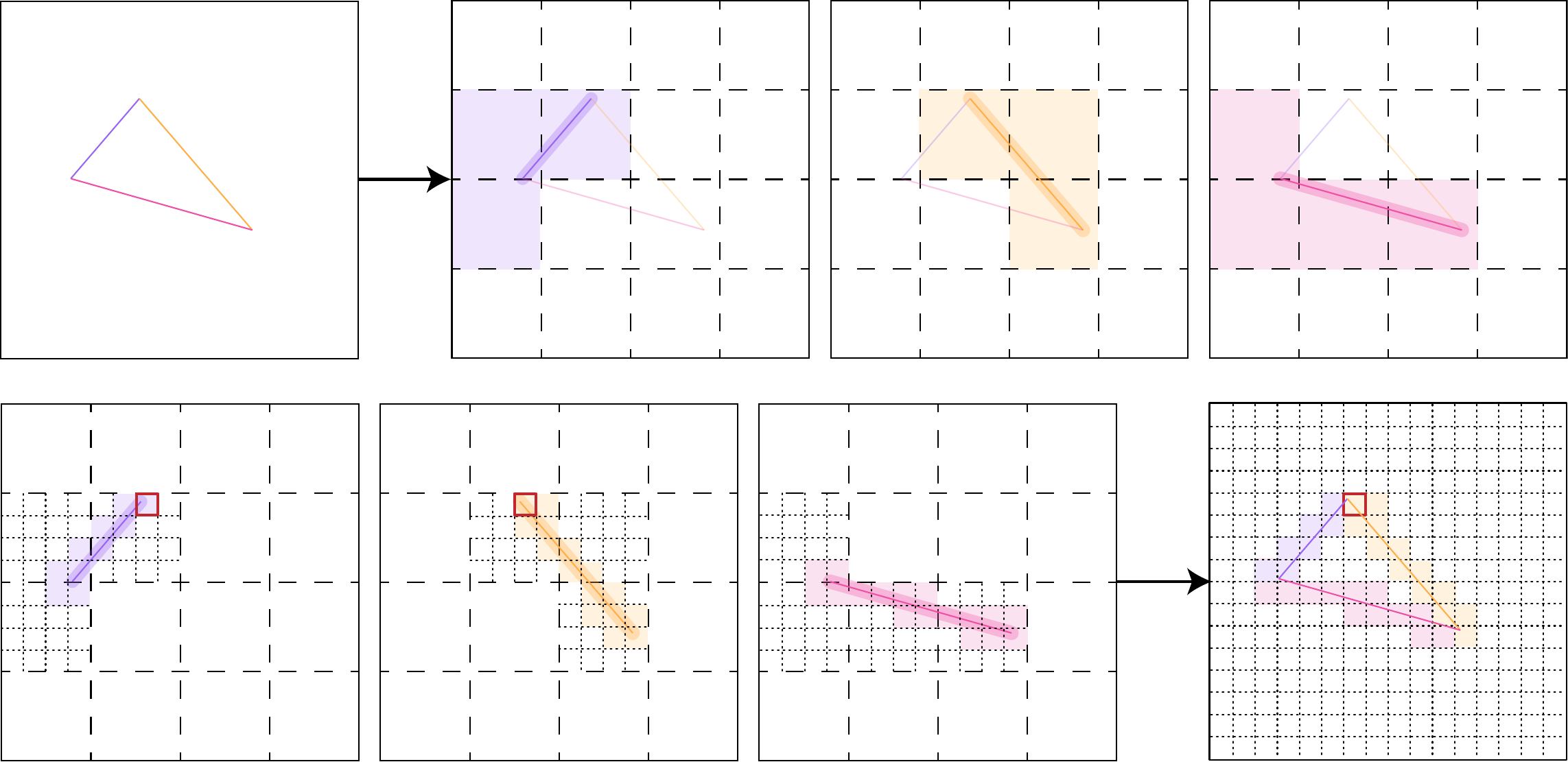}
    \caption{\textbf{Efficient narrow-band UDF computation with Voxel Grid Hierarchy} (2D illustration; line segments represent triangles in 3D). We filter the valid voxel-triangle pairs (denoted by three colors) from the highest level to the lowest. At the lowest level, we compute the distance for all valid voxel-triangle pairs.  For each voxel, we take the minimum distance between it and the remaining triangles as the UDF value.}
    \label{fig:udfcomp}
\end{figure}

\section{PROBLEM DEFINITION AND METHOD OVERVIEW}

We aim to design a GPU algorithm that converts an arbitrary input mesh, $\mathcal M_\mathrm{in}$, into a low-poly approximation, $\mathcal M_\mathrm{out}$, possessing the following properties: (1) $\mathcal M_\mathrm{out}$ is watertight, manifold, and intersection-free; (2) The number of faces in $\mathcal M_\mathrm{out}$ is below a user-specified threshold; (3) $\mathcal M_\mathrm{out}$ closely resembles $\mathcal M_\mathrm{in}$, as measured by Chamfer and Hausdorff distances. 

As illustrated in Figure~\ref{fig:pipeline}, our method tackles this challenge in three stages: \textbf{Parallel Remeshing} converts a mesh in the wild $\mathcal M_\mathrm{in}$ into a manifold, intersection-free mesh $\mathcal{M}_\mathrm{r}$. $\mathcal{M}_\mathrm{r}$ features uniformly sized triangles and edges, enabling faster BVH querying for intersection checks. \textbf{Parallel Mesh Simplification} performs parallel edge collapsing to reduce $\mathcal{M}_\mathrm{r}$ into a low-poly mesh $\mathcal{M}_\mathrm{s}$ with the desired face count. This stage also prevents the introduction of any self-intersections. \textbf{Parallel Safe Projection} moves the surface vertices of $\mathcal{M}_\mathrm{s}$ to mitigate surface shifts and restore sharp features, resulting in the creation of $\mathcal M_\mathrm{out}$. $\mathcal M_\mathrm{out}$ has smaller Chamfer and Hausdorff distances from $\mathcal M_\mathrm{in}$, and maintains the intersection-free property of the mesh. In our implementation, we normalize the vertices of $\mathcal M_\mathrm{in}$ into $[0,1]^3$.

\section{Remeshing}

We first process a mesh in the wild into a signed distance field (SDF) volume, then apply Dual Marching Cubes~\cite{nielson2004dual,wenger2013isosurfaces} on the volume to obtain watertight, manifold, and intersection-free meshes.
All procedures are designed to run efficiently on GPUs.

\subsection{Mesh-to-Volume}

In this stage, we first compute an unsigned distance field (UDF) on voxel grids around the input mesh surface.
UDF is the minimum distance to each voxel center among all triangles. A brute-force computation would require computing the distance between all triangle-voxel pairs.
However, it is easy to reach millions of voxels and triangles in practice.
Fortunately, for remeshing, we only care about the values in the narrow band around the mesh surface where the mesh intersects the voxels. Taking this into consideration, we design an efficient UDF computation process, illustrated in Figure~\ref{fig:udfcomp}. We build a hierarchy of voxel grids of increasing resolutions and filter triangle-voxel pairs outside the narrow band at each level in the hierarchy. At the lowest level in the hierarchy, we perform a brute-force computation of the remaining triangle-voxel pairs to obtain a grid of distance values. For each voxel, we take the minimum of all distances between the voxel and its corresponding remaining triangles, as by definition the UDF value at the voxel.

\begin{figure}[t]
\begin{center}
   \includegraphics[width=\linewidth]{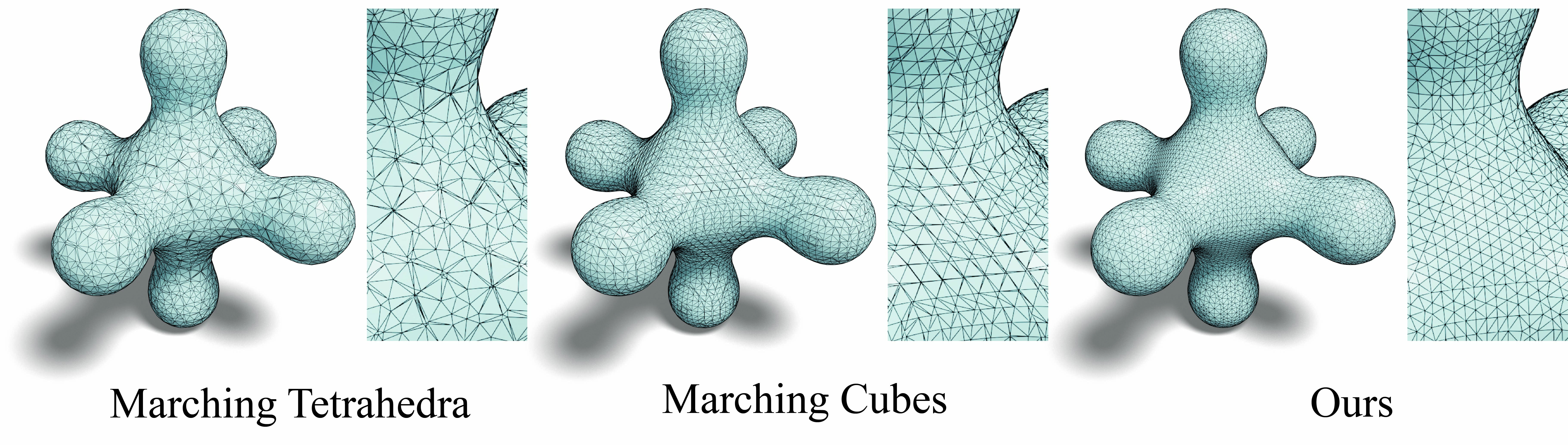}
\end{center}
\vspace{-1.0em}
  \caption{DualMC generates meshes containing much fewer skinny triangles compared with other iso-surface extraction algorithms.}
\vspace{-1.5em}
\label{fig:dualmc_adv}
\end{figure}

Specifically, we filter a triangle-voxel pair away if the voxel does not intersect with a narrow band around the triangle. In our implementation, we set the width of the narrow band to $3/R_\mathrm{DMC}$ (the same value as OpenVDB~\cite{museth2013openvdb}), where $R_\mathrm{DMC}$ is the resolution for DualMC. This is a safe boundary for DualMC since it only considers values on the voxel corners that have a sign change.
However, it is unnecessarily expensive to compute the intersection exactly. In practice, we employ a conservative approximation -- if the bounding sphere of the voxel does not intersect the narrow band around the triangle, the voxel cannot intersect the band either. This can be easily tested by computing a point-triangle distance.
Since input triangles can be trivially divided into batches during the computation, the memory usage does not depend on the complexity of the input mesh. The algorithm does not rely on any topology assumptions from the input, allowing it to operate on triangle soups effectively.

\subsection{Volume-to-Mesh}

\begin{wrapfigure}{l}{0.04\textwidth}
    \centering
    \vspace{-1.5em}\includegraphics[width=0.08\textwidth]{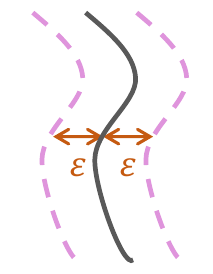}
    \vspace{-2.5em}
\end{wrapfigure}

We take the output of the mesh-to-volume algorithm and extract the watertight, self-intersection-free, manifold mesh from the implicit field. We first turn the UDF generated by the mesh-to-volume algorithm it into SDF by adding a band to the surface. Specifically we subtract a threshold from the UDF values: $s(x) = u(x) - \epsilon$, where $\epsilon$ is the length of the band and requires a value of $\sqrt{3}/ (2 R_\mathrm{DMC})$ to ensure DualMC does not skip any sign changes (for determining the boundaries). In our implementation, we use $0.9 / R_\mathrm{DMC}$, which is the same choice as~\cite{chen2023robust}.

\paragraph{Iso-Surface Extraction}
We have implemented the Dual Marching Cubes (DualMC) algorithm~\cite{nielson2004dual} on GPU for mesh extraction. Compared to other iso-surface extraction algorithms, DualMC can greatly reduce the number of skinny triangles, as shown in Figure~\ref{fig:dualmc_adv}, which is beneficial to the following mesh simplification process and some downstream applications. 

The algorithm takes SDF as input and generates quad meshes, with each face comprising four vertices. From the previous stage, we obtain a dense grid within the bounding space, where each grid vertex stores its corresponding SDF value. The first step in the original DualMC is constructing patches (yellow in Figure~\ref{fig:dualmc}) for each voxel following a modified Marching Cubes look-up table (see Appendix~1). Unlike MC, the DualMC look-up table consolidates certain connected triangles into a single \textit{patch}, facilitating the subsequent quad-creation process. The patch count and connectivity are determined by the signs of the eight vertices of a voxel. 

Instead of directly traversing all voxels and building the patches, we initially examine voxel occupancy using the look-up table. A voxel is considered \textit{empty} if it does not contain any patches. In parallel, we traverse all voxels and record the occupancy, then establish an indexing map from dense voxels to those in use. This indexing optimizes subsequent operations by reducing the time spent traversing empty voxels.

\begin{wrapfigure}{l}{0.4\linewidth}
    \centering
    \vspace{-1em}
    \includegraphics[width=1.2\linewidth]{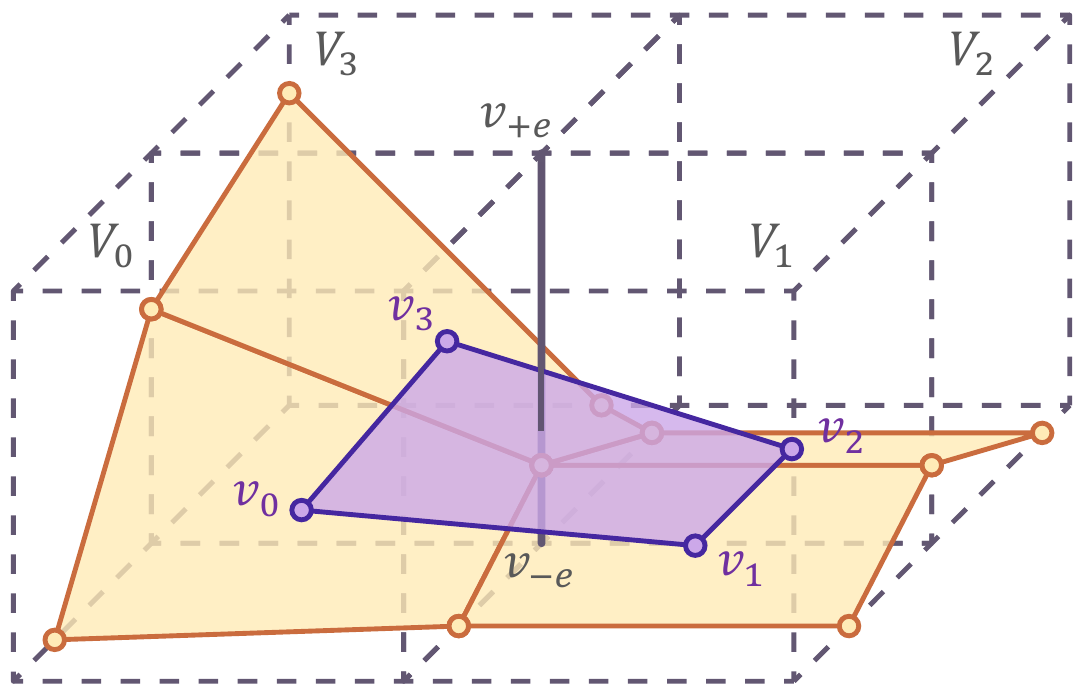}
    \vspace{-1.5em}
    \caption{Patches (in yellow) and a quad (in purple) in DualMC. $(v_{+e}, v_{-e})$ is a valid edge. }
    \vspace{-0.5em}
    \label{fig:dualmc}
\end{wrapfigure}
Next, we proceed to construct patches within each voxel by utilizing the look-up table once more. The vertex positions of these patches are determined through linear interpolation between the grid vertices based on their SDF values. Subsequently, we generate quad faces based on the patches. An edge with different signs at its two ends is said to be \textit{valid}. For each \textit{valid} edge, we identify the neighboring voxels of the edge and connect the center point of the patches to form a quad, as illustrated in Figure~\ref{fig:dualmc}. The quad ($v_0, v_1, v_2, v_3$) on a \textit{valid} edge ($v_{+e}, v_{-e}$) is created from its four neighboring voxels ($V_0, V_1, V_2, V_3$), where $v_{+e}$ has a positive SDF value and $v_{-e}$ has a negative SDF value. 
In voxels containing more than one patch, we consider only the patch with a vertex located on this edge. Throughout the process of building patches and quads, we adhere to 2-kernel \emph{Gather} strategy to ensure algorithmic parallelism. We index non-empty elements in the first kernel, and then write values in the second kernel.

We found that self-intersections can occasionally occur in the results of DualMC, primarily due to improper quad division and numerical instability. We propose two strategies to mitigate these issues and ensure intersection-free results in practice (see Section~\ref{sec:inter_correct}). Specifically, we adopt the concept of the \textit{envelope} from~\cite{ju2006intersection,wang2009intersection} to eliminate intersections caused by quad division and introduce a smoothing function for patch vertex creation to address numerical instability. More implementation details are provided in Appendix~2.1.

\section{Mesh Simplification}

In this section, we propose a parallel intersection-free simplification method (Algorithm~\ref{alg:edge_col}) to reduce the triangle count of the remeshed results. Our approach follows the iterative triangle reduction strategy of QEM-based methods~\cite{garland1997surface}.

Instead of performing one collapse at a time, for each iteration, we divide the mesh into independent regions and perform one collapse for each region in parallel. Specifically, we compute costs for each edge (Line~\ref{alg:compute_edge_cost}) and then propagate the costs to the neighborhood of each edge (Line~\ref{alg:cost_propagation}).
Then, we perform one edge collapse for each independent region in parallel (Line~\ref{alg:collapse_edge}).
After each collapse iteration, we check for self-intersections in the resulting mesh and undo the edge collapse operations that contribute to the intersecting triangles (Line~\ref{alg:intersection_check}-Line~\ref{alg:undo_collapse}). We introduce an efficient and robust triangle intersection detector that can accurately detect both coplanar and three-dimensional triangle intersections.

\begin{algorithm}
\caption{Parallel Intersection-Free Mesh Simplification. $E(\mathcal M)$ denotes the edge set and $F(\mathcal M)$ denotes the face set.}\label{alg:edge_col}
\DontPrintSemicolon
\SetKwComment{Comment}{/* }{ */}
\SetKwFunction{FMain}{ParallelSimplification}
  \SetKwProg{Fn}{Procedure}{:}{}
  \Fn{\FMain{$\mathcal M_\mathrm{1}$, $\mathcal N_\mathrm{t}$}}{
        
      \KwInput{Manifold mesh $\mathcal M_\mathrm{r}$, target \# faces $\mathcal N_\mathrm{t}$}
      \KwOutput{Simplified mesh $\mathcal M_\mathrm{s}$}
  
        $\mathcal M_1 \gets \mathcal M_\mathrm{r}$\;

        \While{$|F(\mathcal M_1)| > \mathcal N_\mathrm{t}$}{
            \ForEach(\tcp*[h]{(parallel) edge cost}){$e \in E(\mathcal M_1)$}
            {
                $C_e \gets \operatorname{ComputeEdgeCost}(\mathcal M_\mathrm{1},e)$ \tcp*[h]{Eq. 1}
                \label{alg:compute_edge_cost}
            }
            \ForEach(\tcp*[h]{(parallel) cost propagation}){$e \in E(\mathcal M_1)$}
            {
                \ForEach{\textup{triangle} $t \in \operatorname{Find1RingNeighbors}(\mathcal M_\mathrm{1},e)$}
                {
                    $C_t \gets \operatorname{Min}(C_t, C_e)$ 
                    \label{alg:cost_propagation}
                }
            }
            $\mathcal M_{\mathrm{s}} \gets \mathcal M_{\mathrm{1}}$\;
            \ForEach(\tcp*[h]{(parallel) edge collapse}){$e \in E(\mathcal M_1)$}
            {
                $T_e \gets \operatorname{Find1RingNeighbors}(\mathcal M_\mathrm{1}, e) $\;
                
                \If{$C_e = C_t \;\forall t \in T_e$}
                { 
                    $\mathcal M_\mathrm{s} \gets \operatorname{CollapseEdge}(\mathcal M_\mathrm{s},e)$\;
                    \label{alg:collapse_edge}
                }
            }
            \Repeat (\tcp*[h]{(parallel) intersection check and undo}){ $T_\mathrm{intsct} = \emptyset$ }{
                $T_\mathrm{intsct} \gets \operatorname{IntersectionCheck}(\mathcal M_\mathrm{s})$ \tcp*[h]{Sec.~\ref{sec:int_check}}
                \label{alg:intersection_check}

                $\mathcal M_\mathrm{s} \gets \operatorname{UndoCollapse}(\mathcal M_\mathrm{1}, \mathcal M_\mathrm{s}, T_\mathrm{intsct})$ \tcp*[h]{Sec. 5.3}
                \label{alg:undo_collapse}
            }
            $\mathcal M_\mathrm{1}\gets\mathcal M_\mathrm{s}$\;
        }
    \KwRet $\mathcal M_\mathrm{s}$\;
  }
\end{algorithm}


\subsection{Parallel Edge Collapse}\label{sec:qem_parallel}

The algorithm works in an iterative procedure. In every iteration, to achieve the parallel edge-collapsing algorithm, we first define an \emph{edge cost} for choosing the best collapse candidates. Then we propagate the edge costs to its neighboring triangles and use this to find independent regions. Next, parallel edge collapsing is conducted within each independent region to reduce triangle counts.

\paragraph{Edge Cost}
We define the edge cost of an edge $e_{ab}$ as follows:
\begin{equation}
\begin{split}
\vspace{-0.3em}
    C_{ab} &= Q_{ab} + w_\mathrm{e} C_\mathrm{e} + w_\mathrm{s} C_\mathrm{s} \\
    C_\mathrm{e} = l_{ab}, \quad C_\mathrm{s} &= \sum_{\Delta_{ijk}} (1-C_{ijk}), \quad C_{ijk} = \frac{4\sqrt{3}A_{ijk}}{l_{ij}^2 + l_{jk}^2+l_{ki}^2}
    \label{eqn:cost}
\vspace{-0.2em}
\end{split}
\end{equation}
where $Q_{ab}$ is the widely used Quadric Error Metrics (QEM)~\cite{garland1997surface}. Additionally, we include the edge length $l_{ab}$ in the cost to prioritize collapsing shorter edges. Similar to ~\cite{liu2020neural}, we include \emph{skinny cost} $C_{ijk}$ defined on each generated triangle $\Delta_{ijk}$ to prevent generating thin triangles, where $A$ is the triangle area and $l$ is the length of a triangle edge. Higher $C_{ijk}$ indicates a more equilateral triangle. Since all of the components in edge cost can be computed locally, the cost computation can be easily parallelized on the GPU. Default values for the weights are set as $ w_\mathrm{e} = 0.001 $ and $ w_\mathrm{s} = 0.005 $.

\paragraph{Cost Propagation}

\begin{figure}
    \centering
    \includegraphics[width=0.7\linewidth]{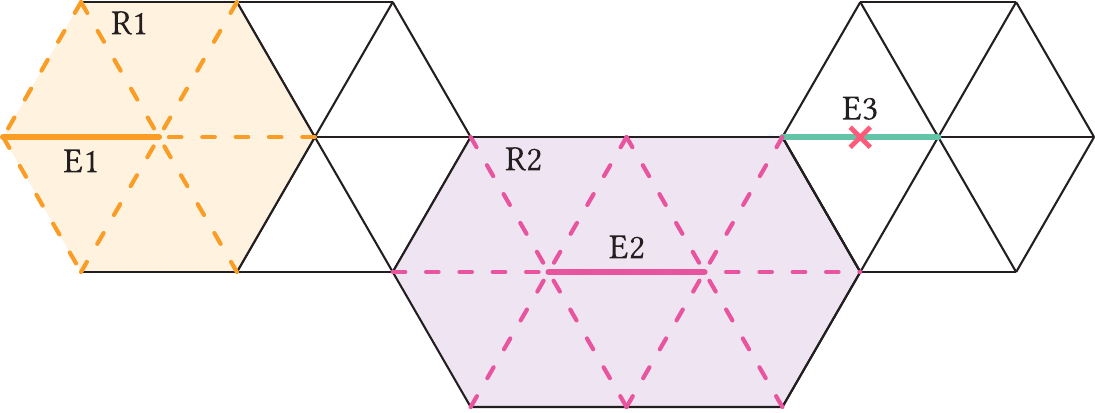}
    \caption{\textbf{Independent regions.} The yellow and pink areas mark two independent regions ($R_1$ and $R_2$) associated with $E_1$ and $E_2$, respectively. $E_3$ does not have an associated independent region since it has neighboring triangles that do not share the same cost with it. }
    \label{fig:local_area}
    \vspace{-1em}
\end{figure}
After computing the cost for each edge, we propagate this cost to all of its 1-ring neighboring triangles. Each triangle then adopts the minimal edge cost as its \emph{triangle cost}. The 1-ring neighboring triangles of an edge are defined as the union set of the 1-ring neighboring triangles of the two vertices of the edge. Next, we identify independent regions by comparing the cost of an edge to the costs of all triangles within its 1-ring neighborhood, following the method described in~\cite{gautron2023interactive}. If \emph{all} the 1-ring neighboring triangles share the exact same cost as the edge, the region consisting of these triangles is defined as an \emph{independent region} associated with the edge (see Figure~\ref{fig:local_area}). 
Since independent regions are determined solely by edge costs, it is crucial to properly distinguish between edges with identical costs to prevent incorrect region division. To achieve this, we use a 64-bit integer to store each edge cost, where the higher-order 32 bits hold the converted edge costs from floats, and the lower-order 32 bits hold a unique edge ID. This configuration allows for the use of the GPU \emph{AtomicMin} operation to compute the minimal edge cost for each triangle effectively and can also eliminate the errors caused by floating-point number comparisons. 


\paragraph{Edge Collapse} After finding the independent regions, edges that are associated with an independent region (e.g. $E_1$ and $E_2$ in Figure~\ref{fig:local_area}) are collapsed in parallel without interfering with each other, while the other edges (e.g. $E_3$) are kept unchanged. The link condition \cite{dey1999topology} is applied to ensure that the manifold property of the mesh is preserved during the edge collapse operations.

The parallel edge-collapse operation is conducted iteratively until the termination criteria are met. In our algorithm, we can either terminate the simplification process by a target edge cost or a target triangle count. In our evaluation experiments, we use the latter to keep consistency with previous algorithms.

\begin{figure}[t]
\centering
   \includegraphics[width=\linewidth]{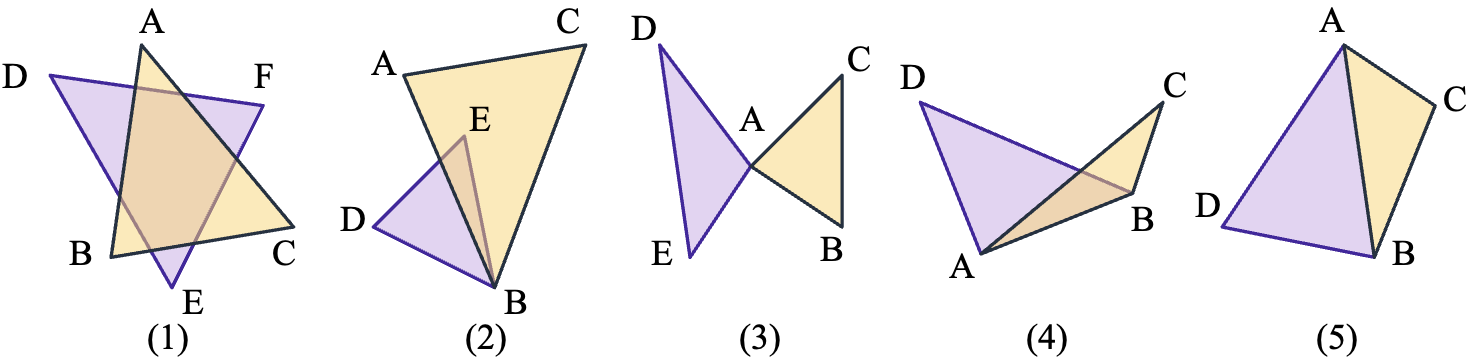}
   \vspace{-2em}
  \caption{\textbf{Coplanar self-intersection cases.} We have to distinguish between self-intersection (1)(2)(4) and vertex/edge sharing between neighboring triangles (3) and (5).}
\label{fig:insect_cases_2D}
\end{figure}

\subsection{Parallel Self-Intersection Detection}\label{sec:int_check}

After one step of edge collapse, we conduct an intersection check between each pair of triangles in the decimated mesh. We begin by employing the parallel LBVH tree algorithm~\cite{karras2012maximizing} to construct a BVH for the triangle mesh. We then test each triangle against the BVH tree to find candidate triangles for potential intersections. The BVH tree is widely used in geometry processing algorithms for acceleration purposes, and we have experimentally found that BVH is crucial for fast intersection tests. While brute-force pairwise intersection tests can take up to one minute on the GPU for meshes with 2 million faces, BVH construction and querying take roughly 10 milliseconds in our setup. 

To accurately determine intersections from BVH queries, we have implemented a robust triangle-triangle intersection test that categorizes intersections into two cases, coplanar triangles and three-dimensional triangles. 
The latter is effectively handled by \textit{Devillers' algorithm}~\cite{devillers2002faster}, thanks to its robustness and efficiency. 


However, for coplanar cases, Devillers' algorithm can fail to differentiate between true self-intersections (Figure~\ref{fig:insect_cases_2D} (2)(4)) and legitimate vertex- and edge-sharing scenarios (Figure~\ref{fig:insect_cases_2D} (3)(5)). To overcome these challenges, our method categorizes triangle pairs based on the number of shared vertices and implements specialized detection algorithms for each category. 
For single-vertex-sharing triangles, we test whether the two internal angles on that vertex overlap. In edge-sharing cases, we test whether the triangles are on the same side of the shared edge. Further details are provided in Appendix~3.

Our self-intersection detection algorithm performs all operations on the GPU, and adopts a mixed-precision approach to achieve the best precision with good performance. Using single precision reduces memory consumption and accelerates computation, yet it can not always meet the precision requirements to correctly detect intersections (see Table~\ref{tab:selfintersect_quant}). Conversely, using double precision for all operations significantly increases GPU memory usage and decreases computational efficiency, which prevents applications on large meshes. To strategically balance precision and performance, we selectively apply double precision only to the most critical operations --- coplanar detection and the computation of intersecting lines between triangles in Devillers' algorithm. This approach effectively prevents false negatives without excessively burdening system resources.

\subsection{Undo Operation}\label{sec:undo}
The self-intersection test produces a list of intersecting triangle pairs and their corresponding collapse operations. We revert those collapse operations by restoring the original edge. 
We call each reversion an \emph{undo operation}.


Due to the parallel nature of edge collapsing, self-intersections may still occur after one iteration of undo operations, as the undo action in one local region can lead to intersections with other remaining parts that are also changed by edge collapsing. To address this, we conduct a self-intersection check after every iteration of undo operations. Multiple iterations of self-intersection testing and edge undoing are performed until no intersections are detected. In a hypothetical worst case, the number of undo iterations could equal the total number of edge collapse operations. However, in our experiments, over 99.99\% of cases are resolved within two undo iterations, with no instances exceeding three (see Table~\ref{tab:worst}).

If the collapse-undo actions are naively repeated each iteration, it is possible that the same edge (with a low cost) can be collapsed and then undone in every iteration, greatly diminishing the algorithm efficiency. To prevent this, invalid flags are introduced for edges that led to self-intersections in the previous iteration, ensuring that they are not considered for collapsing again in the very next iteration.

As the iterations increase and the number of independent regions decreases, it becomes more likely that all collapsed edges will be undone. In such cases, flagging invalid edges for only one iteration is not sufficient to enhance the algorithm efficiency. To address this, we design a strategy where, if no edges are collapsed (i.e., all are undone) in one iteration, we retain the invalid flags until some edges are successfully collapsed or a maximum \textit{tolerance} iteration number (default = 4) is reached.

\section{Safe Projection}

The safe projection stage takes the simplified mesh $\mathcal M_\mathrm{s}$ as input and projects its vertices back towards the original mesh $\mathcal M_\mathrm{in}$ \textit{safely} without causing self-intersection, resulting in an output mesh $\mathcal M_\mathrm{out}$. In addition to reducing the distance between $\mathcal M_\mathrm{out}$ and $\mathcal M_\mathrm{in}$, the safe projection stage also keeps the shapes of triangles in $\mathcal M_\mathrm{out}$ close to those in $\mathcal M_\mathrm{s}$ to preserve the high mesh quality of $\mathcal M_\mathrm{s}$ and prevent the re-emergence of poorly shaped triangles and large bending angles. To achieve these two goals, we first formulate an energy function of any mesh $\mathcal S$ that has the same vertex connectivity as $\mathcal M_\mathrm{s}$. This energy penalizes (1) the distance between $\mathcal S$ and $\mathcal M_\mathrm{in}$ and (2) the amount of deformation from $\mathcal M_\mathrm{s}$ to $\mathcal S$. 
The energy function is then optimized using our collision-aware Newton-type solver, inspired by Incremental Potential Contact (IPC)~\cite{Li2020IPC}. It optimizes the energy function iteratively, deforming the mesh along a piecewise-linear trajectory $\mathcal M_\mathrm{s} = \mathcal S^{0} \rightarrow \mathcal S^{1} \rightarrow \cdots \rightarrow \mathcal S^{T} = \mathcal M_\mathrm{out}$, and guarantees that no self-intersection will occur anywhere on this trajectory. Note that in the safe projection stage, only the positions of mesh vertices are modified, and the connectivity of the output $\mathcal M_\mathrm{out}$ stays the same as the input $\mathcal M_\mathrm{s}$.

\subsection{Energy Definition}  \label{sec:energy}

In this section, we use $x_i \in \mathbb R^3$ to denote the position of the $i$-th vertex of a mesh $\mathcal S$, and denote the concatenated vector of $\{x_i\}$ as $X = (x_1^\top, x_2^\top, \cdots, x_{n}^\top)^\top\in \mathbb R^{3n}$, where $n$ is the number of vertices.
We use $\mathcal S(X)$ to denote the mesh with the same connectivity as $\mathcal M_\mathrm{s} = \mathcal S(X^{0})$ but the vertex positions moved to $X$. Let $V(\mathcal S)$ denote the set of vertices, $F(\mathcal S)$ denote the set of triangular faces, and $E(\mathcal S)$ denote the set of edges of mesh $\mathcal S$. 

\paragraph{The distance energy} To penalize the distance between $\mathcal S$ and $\mathcal M_\mathrm{in}$, we construct a distance energy function $E_\mathrm{dis} = E_\mathrm{S2M} + E_\mathrm{M2S}$ that approximates the mesh Chamfer Distance
\begin{align}
\begin{split}
    CD(\mathcal S, \mathcal M_\mathrm{in}) = \iint_{\mathcal S} \min_{y \in \mathcal M_\mathrm{in}} \|x - y\|^2 \dd s(x) \\
    + \iint_{\mathcal M_\mathrm{in}} \min_{x \in \mathcal S} \|y - x\|^2 \dd s(y).  
\end{split}
\label{eqn:chamfer}
\end{align}

We approximate the first integral in Equation \ref{eqn:chamfer} by discretizing the surface area into Voronoi areas associated with each vertex of $\mathcal S$. Those Voronoi areas are further approximated by their counterparts on $\mathcal S^0$, resulting in the following approximated distance:
\begin{align}
    E_\mathrm{S2M}(X) := \sum_{i \in V(\mathcal S)} s^0_i \min_{y \in \mathcal M_\mathrm{in}} \|x_i - y\|^2, \label{eqn:E_S2M}
\end{align}
where $s^0_i$ is the Voronoi area of vertex $i$ on $\mathcal S^0$. This approximation enables efficient computation and optimization of the distance. 

To discretize the second integral in Equation \ref{eqn:chamfer}, we avoid using $V(\mathcal M_\mathrm{in})$ as in Equation \ref{eqn:E_S2M} because the vertices are often too numerous and unevenly distributed across the surface of $\mathcal M_\mathrm{in}$. Instead, we uniformly sample $m$ points, denoted by ${y_1, y_2, \ldots, y_m}$, across the surface of $\mathcal M_\mathrm{in}$. We then define 
\begin{align}
    E_\mathrm{M2S}(X) := \frac{A(\mathcal M_\mathrm{in})}{m} \sum_{j=1}^{m} \min_{x \in \mathcal S(X)} \|y_j - x\|^2, \label{eqn:E_M2S}
\end{align}
where $A(\mathcal M_\mathrm{in})$ is the surface area of mesh $\mathcal M_\mathrm{in}$. In our experiments, we fix $m=16384$. 

Although minimizing the distance energy $E_\mathrm{dis} = E_\mathrm{S2M} + E_\mathrm{M2S}$ can pull mesh $\mathcal S$ towards $\mathcal M_\mathrm{in}$, it can also distort the triangular faces of $\mathcal S$ and the dihedral angles between them, potentially leading to the re-emergence of oversized, undersized, or skinny triangles and undesirably large bending angles (Figure~\ref{fig:ablation_energy}). To address this, we take cues from the Finite Element Method (FEM) and model the mesh as a ``hyperelastic shell'' \cite{grinspun2003discrete, tamstorf_discrete_2013} to keep the shape of $\mathcal S$ close to that of $\mathcal M_\mathrm{s}$. The deformation energy of a hyperelastic shell consists of two terms: the (membrane) elastic energy $E_\mathrm{elas}$ measures the amount of stretching and shearing of the triangular faces, and the bending energy $E_\mathrm{bend}$ measures the amount of change in the dihedral angles between the triangular faces. 

\paragraph{The elastic energy} An elastic energy of a triangular face should depend solely on its deformation within its supporting plane. It is also desired for the elastic energy to be rotationally invariant, penalizing only stretching and shearing deformations. Drawing from continuum mechanics, we employ the simplest elasticity model, the St. Venant-Kirchhoff model \cite{10.1145/2343483.2343501}, which meets all these requirements. We define our elastic energy as
\begin{align}
    E_\mathrm{elas}(X) := \sum_{\triangle_{ijk} \in F(\mathcal S)} \frac14 A^0(\triangle_{ijk}) \|\mathbf F_{ijk}^\top \mathbf F_{ijk} - \mathbf I\|_F, \label{eqn:E_elas}
\end{align}
where $\|\cdot\|_F$ denotes the Frobenius norm, $A^0(\triangle_{ijk})$ is the area of triangle $(x^0_i, x^0_j, x^0_k)$, and $\mathbf F_{ijk} = \mathbf F_{ijk}(X, X^0) \in \mathbb R^{2\times 2}$ is the \textit{deformation gradient} of the triangle. The deformation gradient $\mathbf F_{ijk}$ is determined by solving the linear equations
\begin{align}
    \mathbf P_{ijk}\begin{bmatrix}
        x_j - x_i, x_k - x_i
    \end{bmatrix} = \mathbf F_{ijk} \mathbf P^0_{ijk}\begin{bmatrix}
        x^0_{j} - x^0_{i}, x^0_{k} - x^0_{i}
    \end{bmatrix},
\end{align}
where $\mathbf P_{ijk} \in \mathbb R^{2\times 3}$ transforms points in the three-dimensional coordinate system to an arbitrary two-dimensional coordinate system of the supporting plane of points $x_i, x_j$ and $x_k$. The matrix $\mathbf P^0_{ijk}$ is defined similarly with respect to the initial configuration $X^0$. Equation \ref{eqn:E_elas} is equivalent to a St. Venant-Kirchhoff elastic energy with Young's modulus $E = 1$ and Poisson's ratio $\nu = 0$.

\paragraph{The bending energy} We define the bending energy as 
\begin{align}
    E_\mathrm{bend}(X) := \sum_{(i,j) \in E(\mathcal S)} \frac12 \|x^0_{i} - x^0_{j}\| (\theta_{ij} - \theta^0_{ij})^2,  \label{eqn:E_bend}
\end{align}
where $\theta_{ij} = \theta_{ij}(X)$ is the dihedral angle between the two triangular faces sharing edge $(i, j)$. 

While the motivation of introducing $E_\mathrm{elas}$ and $E_\mathrm{bend}$ is to enhance the shape quality of the output mesh, empirically they also help avoid local minima and reduce the distance between $\mathcal S$ and $\mathcal M_\mathrm{in}$ (see Section~\ref{sec:analysis_stage3}). Combining all the energies defined above, we have the total energy 
\begin{align}
    E = k_\mathrm{dis} E_\mathrm{dis} + k_\mathrm{elas} E_\mathrm{elas} + k_\mathrm {bend} E_\mathrm{bend}.
\end{align}
In our experiments, we set $k_\mathrm{dis}=10^3, k_\mathrm{elas}=10^{-1}, k_\mathrm{bend}=10^{-2}$.

\subsection{Self-Intersection-Free Optimization}

\begin{algorithm}
\caption{Safe Projection}\label{alg:ipc}
\DontPrintSemicolon
\SetKwComment{Comment}{/* }{ */}
\SetKwFunction{FMain}{SafeProjection}
  \SetKwProg{Fn}{Procedure}{:}{}
  \Fn{\FMain{$\mathcal M_\mathrm{s}$, $\mathcal M_\mathrm{in}$, $\{k\}$, $\hat d$, $T$}}{
        \KwIn{\begin{itemize}
            \item[] $\mathcal M_\mathrm{s}$: the simplified mesh from stage 2
            \item[] $\mathcal M_\mathrm{in}$: the original input mesh
            \item[] $\{k\}$: coefficients of each energy component
            \item[] $\hat d$: distance threshold for the barrier function $b$
            \item[] $T$: number of iterations
        \end{itemize}
        }
        \KwOut{$\mathcal M_\mathrm{out}$: the projected mesh}

        

        $X^0 \gets$ vertex positions of $\mathcal M_\mathrm{s}$ \;
        $t \gets 0$\;
        \While{$t < T$}{
            $\mathbf H \gets \operatorname{SPDProject}\left( \nabla^2 B \left(X^t\right) \right)$\; \label{alg:spdproject}
            Solve $p$ from $\mathbf H p + \nabla B\left(X^t\right) = 0$\; \label{alg:linear_solver}
            $\alpha \gets \min\left(1, \operatorname{ACCD}\left(X^t, p\right)\right)$\; \label{alg:ccd}
            \Repeat{$B(X) < B(X^t)$}{ \label{alg:line_search_begin}
              $X \gets X^t + \alpha p$\;
              $\alpha \gets \alpha / 2$\;
            } \label{alg:line_search_end}
          $t \gets t + 1$\;
        }
        $\mathcal M_\mathrm{out} \gets \mathcal S(X^T)$\;
        \KwRet $\mathcal M_\mathrm{out}$\;
  }
\end{algorithm}

We present a customized Newton-type algorithm (Algorithm \ref{alg:ipc}) to optimize the energies defined in Section \ref{sec:energy} iteratively while enforcing the self-intersection-free guarantee. 

At the $t$-th iteration, we obtain the new vertex positions $X^{t+1}$ through optimizing
\begin{equation}
    X^{t+1} = \underset{X}{\operatorname{argmin ~}} B(X),
\end{equation}
where the \textit{augmented energy} function $B(x)$ is defined, following IPC \cite{Li2020IPC}, as
\begin{equation}
    B(X) = E(X) + k_\mathrm{bar} \sum_{c \in C} b(d_c(X)),  \label{eqn:augmented_energy}
\end{equation}
where $C$ are all possible \textit{contact pairs}, including point-triangle pairs and edge-edge pairs on mesh $\mathcal S$, and $d_c$ is the distance between the pair $c$. To enforce the self-intersection-free constraint, contact pairs closer than $\hat d$ are penalized by the \textit{barrier function} 
\begin{align}
    b(d) = \begin{cases}
        -(d-\hat d)^2 \ln (d/\hat d), & 0 < d < \hat d, \\
        0, &  d \ge \hat d,
    \end{cases}
\end{align}
which goes to infinity as $d$ approaches zero. We set $k_\mathrm{bar} = 10^2$ and $\hat d = 10^{-3}$ in our experiments. 

We employ the projected Newton's method to optimize $B(X)$. At the $t$-th iteration, the Hessian matrix $\nabla^2 B(X^t)$ is projected to a nearby Symmetric Positive Definite (SPD) matrix $\mathbf H$ by eigenvalue modification~\cite{nocedal_numerical_2006} (Line \ref{alg:spdproject}). We solve for the update direction $p$ using the Conjugate Gradient method (Line \ref{alg:linear_solver}). We provide details about gradient and Hessian computation in Appendix~2.2. To guarantee no self-intersection anywhere on the linear trajectory $X^t \rightarrow X^{t+1} = X^t+ \alpha p$, we first compute a safe, intersection-free step size $\alpha$ along $p$ using Additive Continuous Collision Detection (ACCD)~\cite{Li2021CIPC} (Line \ref{alg:ccd}). A line search is then conducted to find a step size in $(0, \alpha]$ that decreases the energy $B(X^{t+1})$ from $B(X^t)$ (Line \ref{alg:line_search_begin}-\ref{alg:line_search_end}). As a result, mesh $\mathcal S$ deforms along a piecewise linear trajectory $X^0 \rightarrow X^1 \rightarrow \cdots \rightarrow X^{T}$ to reduce the augmented energy $B$, where every linear piece does not introduce self-intersection. 


In our experiments, we run $T = 50$ iterations and observed satisfactory quality of output meshes (see Section~\ref{sec:analysis_stage3}). To reduce the cost of nearest neighbor finding in Equation~\ref{eqn:E_S2M} and \ref{eqn:E_M2S}, we update the target positions every 10 iterations (see Appendix~2.2 for details on their computation).


\begin{table*}[htbp]
\small
\centering
\caption{Quantitative comparison with baseline methods, with $1\%$ decimation ratio. Our method has a one hundred percent successful rate ($R_s$) on arbitrary mesh inputs and consistently produces manifold ($R_m$) and intersection-free ($R_i$) surfaces with larger minimal triangle angles ($M_t$), lower Hausdorff distance (HD) and Chamfer distance (CD) from input mesh, while being magnitudes faster. {\footnotesize *~\cite{gautron2023interactive} is performed to further simplify meshes. 
}
}
\begin{tabular}{lccccccccccc}
\toprule
\multirow{2}*{Method} & \multirow{2}*{Time$\downarrow$ (s)} & \multirow{2}*{\#V} & \multirow{2}*{\#F} & \multicolumn{2}{c}{HD$\downarrow$ ($10^{-2}$)} & \multicolumn{2}{c}{CD$\downarrow$ ($10^{-4}$)} & \multirow{2}*{$M_t$$\uparrow$ (deg)} & \multirow{2}*{$R_m$$\uparrow$} & \multirow{2}*{$R_i$$\uparrow$} & \multirow{2}*{$R_s$$\uparrow$}\\
\cmidrule(lr){5-6}
\cmidrule(lr){7-8}
 & & & & avg & std & avg & std & & & &\\
\midrule
QEM~\cite{garland1997surface} &115.07 &3882 &7304 &3.375 & 1.182& 4.035& 2.045& 0.265& 56\%&12\% &97\% \\
Progressive Meshes~\cite{hoppe1996progressive} &31.90 &4207 &7300 &5.318 & 10.34& 25.57& 147.8& 1.962& 57\%&22\% &\textbf{100\%} \\
Blender~\cite{blender} &6.83 &3734 &7789 &\textbf{2.054} & 5.450& 3.202& 1.914& 0.277&59\% &25\% & \textbf{100\%}\\
Manifold*~\cite{huang2018robust} &4.73 & 3499&7141 &3.701 & 4.580 &0.9798 & 2.421 & 3.383 &\textbf{100\%} &64\% &\textbf{100\%} \\
ManifoldPlus*~\cite{huang2020manifoldplus} & 29.23&3792& 8928&4.165 & 4.826& 0.8208& 2.237&0.000 &0\% & 0\%&46\% \\
TetWild*~\cite{hu2018tetrahedral} & 419.70& 3326& 6827& 4.698& 9.207& 14.42& 94.64&3.852 & 68\%& 59\%& \textbf{100\%}\\
fTetWild*~\cite{hu2020fast} &275.82 & 3270 &6708 & 6.222& 23.20& 185.5& 147.1&\textbf{4.453} &97\% &69\% & \textbf{100\%}\\
AlphaWrapping*~\cite{portaneri2022alpha} &56.21 & 3521 & 7157&3.369 &4.617 & 0.6451&2.248 &4.081 & \textbf{100\%}&85\% &\textbf{100\%} \\
RoLoPM~\cite{chen2023robust} & 190.08 & 3605 & 7223 & 3.541 & 3.479 & 0.7162 & 2.333 & 0.018 & \textbf{100\%} & \textbf{100\%} & \textbf{100\%} \\
Gautron et al.~\cite{gautron2023interactive} &2.80 &3649 &7517 &4.318 & 5.302 & 2.128 & 6.885 &0.533 & 58\%& 20\% & 99\% \\
\midrule
Ours (w/o stage 3) & \textbf{0.86} & 3536 & 7186 & 3.439 & 4.636 & 0.8468 & 2.416  & 3.317  & \textbf{100\%} &\textbf{100\%} &\textbf{100\%} \\
Ours          &1.95  & 3536 & 7186 & 2.294 &  3.493&  \textbf{0.1846}& 0.4621 &2.994 & \textbf{100\%} & \textbf{100\%} & \textbf{100\%} \\
\bottomrule
\end{tabular}
\label{tab:baseline_comp}
\end{table*}

\begin{table*}[htbp]
\small
\centering
\caption{Quantitative comparison with baseline methods on the same dataset as Table \ref{tab:baseline_comp} but using larger decimation ratios.
}

\begin{tabular}{clccccccccc}
\toprule
\multirow{2}*{Dec. Ratio} & \multirow{2}*{Method} & \multirow{2}*{Time$\downarrow$ (s)} & \multicolumn{2}{c}{HD$\downarrow$ ($10^{-2}$)} & \multicolumn{2}{c}{CD$\downarrow$ ($10^{-4}$)} & \multirow{2}*{$M_t$$\uparrow$ (deg)} & \multirow{2}*{$R_m$$\uparrow$} & \multirow{2}*{$R_i$$\uparrow$} & \multirow{2}*{$R_s$$\uparrow$}\\
\cmidrule(lr){4-5}
\cmidrule(lr){6-7}
& & &avg & std & avg & std & & & &\\
 \midrule
 \multirow{5}* {10\%} & QEM~\cite{garland1997surface} &109.74 & \textbf{1.233}& 3.418& 1.186& 8.624& 0.103& 55\%&4\% &98\% \\
& Gautron et al.~\cite{gautron2023interactive} &1.00 &1.920 &4.124 &1.230 &5.788 &0.501 & 56\%& 19\% &\textbf{100\%} \\
& RoLoPM~\cite{chen2023robust} & 327.20 & 3.525 & 3.495 & 0.6901 & 2.331 & 0.003 & \textbf{100\%} &\textbf{100\%} &\textbf{100\%} \\
\cmidrule[0.2pt](lr){2-11}
& Ours (w/o stage 3) & \textbf{0.70} &2.777 &4.464 & 0.6968 & 2.206 &3.292 &\textbf{100\%} &\textbf{100\%} & \textbf{100\%}\\
& Ours          &3.01 & 1.886&  3.824& \textbf{0.1407}& 0.6642 &4.486 & \textbf{100\%} &\textbf{100\%} &\textbf{100\%} \\
 \midrule
\multirow{5}* {20\%} & QEM~\cite{garland1997surface} &94.98 &\textbf{0.920} &2.258 & 0.3161& 2.046& 0.100&55\% &4\% &99\% \\
& Gautron et al.~\cite{gautron2023interactive} &0.80 & 1.322& 3.378 & 0.7949 &3.661 &0.392 &55\%  &19\% & \textbf{100\%}\\
& RoLoPM~\cite{chen2023robust} & 476.65 & 3.532 & 3.497 & 0.6885 & 2.334 & 0.001 & \textbf{100\%} &\textbf{100\%} &\textbf{100\%} \\
\cmidrule[0.2pt](lr){2-11}
& Ours (w/o stage 3) & \textbf{0.64}  & 2.733 & 4.470 & 0.6943 & 2.200 &  2.938 & \textbf{100\%} &\textbf{100\%} &\textbf{100\%} \\
& Ours          & 4.30 & 1.910 & 3.862&  \textbf{0.1631}& 0.7285& 5.301&\textbf{100\%} & \textbf{100\%}& \textbf{100\%}\\
\bottomrule
\end{tabular}

\label{tab:larger_ratio}
\end{table*}

\begin{figure*}[htbp]
\centering
\includegraphics[width=\linewidth]{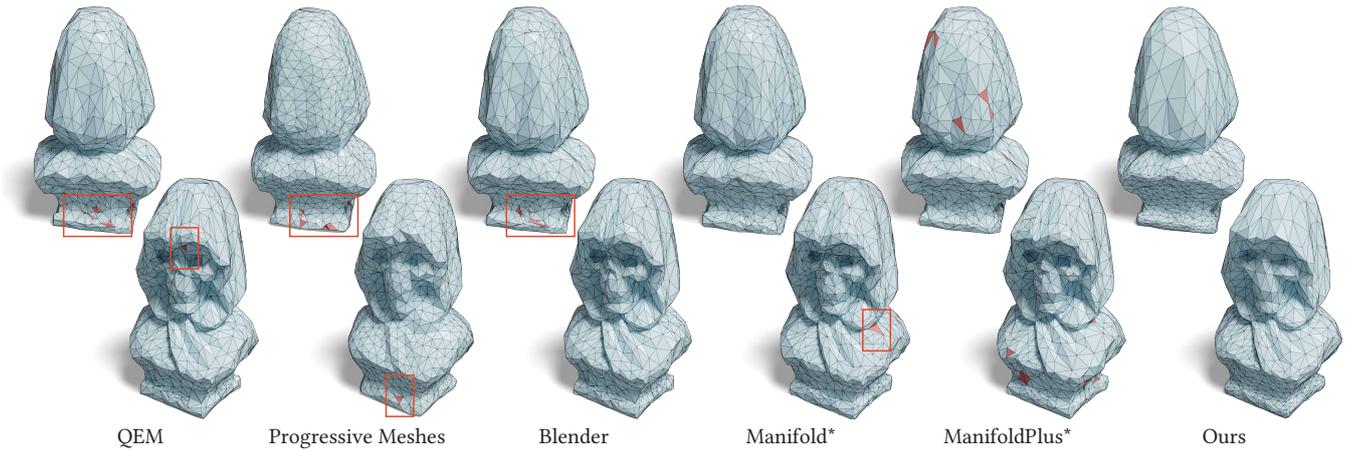}
\vspace{-2.0em}
\caption{\textbf{Qualitative comparison with baseline methods.} We visualize both the front and back of an object. Our method guarantees intersection-free output meshes, whereas baseline models cannot avoid producing self-intersecting triangles (marked in red). }
\label{fig:baseline_inter}
\end{figure*}

\begin{figure*}[t]
\begin{center}
  \includegraphics[width=\linewidth]{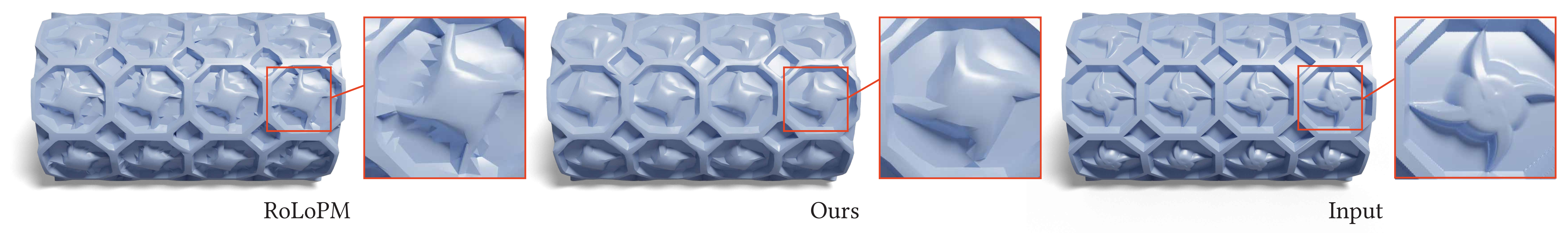}
\end{center}
\vspace{-1.0em}
 \caption{\textbf{Qualitative comparison between Ours and RoLoPM.} The original shape is preserved better in our output mesh, with fewer spike artifacts.}
\label{fig:rolopm}
\end{figure*}

\section{Results}

We present experimental results for our method, including comparisons with baselines and analysis of individual modules. All metrics are collected on a Linux workstation with 1 NVIDIA RTX 4090 GPU and Intel i9-12900K CPU with 24 logical threads except RoLoPM, whose released executable is executed on a Windows server with similar specs.


We compare the low-poly modeling performance between our method and simplification methods QEM~\cite{garland1997surface}, Progressive Meshes~\cite{hoppe1996progressive}, Blender decimation~\cite{blender}, RoLoPM~\cite{chen2023robust}, Gautron et al.~\cite{gautron2023interactive} and remeshing methods Manifold~\cite{huang2018robust}, ManifoldPlus~\cite{huang2020manifoldplus}, TetWild~\cite{hu2018tetrahedral}, fTetwild~\cite{hu2020fast}, AlphaWrapping~\cite{portaneri2022alpha}. For remeshing methods, we further simplify the results using Gautron et al.~\cite{gautron2023interactive}. Note that only RoLoPM can preserve all the mesh properties like our method. We use the implementation from the Wildmeshing-Toolkit~\cite{jiang2022declarative} for QEM and Progressive Meshes. We implement Gautron et al.~\cite{gautron2023interactive} on GPU by ourselves since it has no public code and revise its edge cost into ours.


The comparison is conducted on the Thingi10K dataset~\cite{Thingi10K}. Due to the time-consuming nature of many baseline methods, which can take minutes to process a single mesh, we perform our comparisons on a subset of 100 meshes. These were randomly selected from all meshes in the dataset that have more than 300k faces, and include 43 non-manifold meshes and 76 self-intersecting meshes. The average number of faces is 722,472 and the average number of vertices is 361,632. All methods tested are required to reduce the number of faces to 1\% of the original mesh's count. 

We evaluate all methods using the following metrics: Hausdroff distance (HD), Chamfer distance (CD), and the minimal internal angle among all triangles ($M_t$). For each metric, we report the mean value over the 100-mesh subset. Additionally, we report the rate of each method producing manifold meshes ($R_m$), intersection-free meshes ($R_i$), and the rate of the algorithm finishing without errors ($R_s$). Quantitative results are summarized in Table ~\ref{tab:baseline_comp}, and visual comparisons are provided in Figure~\ref{fig:baseline_inter} and~\ref{fig:rolopm}. 

We also evaluate our method on the \emph{entire} Thingi10K dataset (excluding 5 corrupted meshes), again using a 1\% face decimation ratio, 
and summarize the results in Table~\ref{tab:full_eval}. 
To compute the signed distance field, we use a grid resolution of $R = 256$ by default, $R = 128$ for target faces fewer than 1k, and $R = 64$ for target faces fewer than 50.  

\setlength{\tabcolsep}{2pt}
\begin{table}[htbp]
\small
\centering
\caption{Evaluation results on the \emph{entire} Thingi10K dataset with the decimation ratio set to 1\%.}

\begin{tabular}{lcccccccc}
\toprule
\multirow{2}*{Method} & \multirow{2}*{Time (s)} & \multicolumn{2}{c}{HD ($10^{-2}$)} & \multicolumn{2}{c}{CD ($10^{-4}$)} & \multirow{2}*{$M_t$ (deg)}& \multirow{2}*{$R_i$}\\
\cmidrule(lr){3-4}
\cmidrule(lr){5-6}
 & & avg & std & avg & std &\\
\midrule
Ours (w/o stage 3) &0.40  &4.553  & 3.572& 2.843&7.133 & 3.027&100\%\\
Ours          & 1.23& 2.907 & 2.325 & 0.7116 &3.739 &1.638&100\%\\
\bottomrule
\end{tabular}
\label{tab:full_eval}
\end{table}


While our method is primarily designed for low-poly mesh modeling, we demonstrate in Table~\ref{tab:larger_ratio} that it also performs well with larger decimation ratios. On the 100-mesh subset with 10$\%$ and 20$\%$ decimation ratios, our method achieves CD and HD distances comparable to the state-of-the-art mesh decimation methods, while maintaining better mesh properties and fast processing speeds.



\section{Ablation Studies}

\subsection{Overall Performance}

\begin{figure}[htbp]
\begin{center}
   \includegraphics[width=1.0\linewidth]{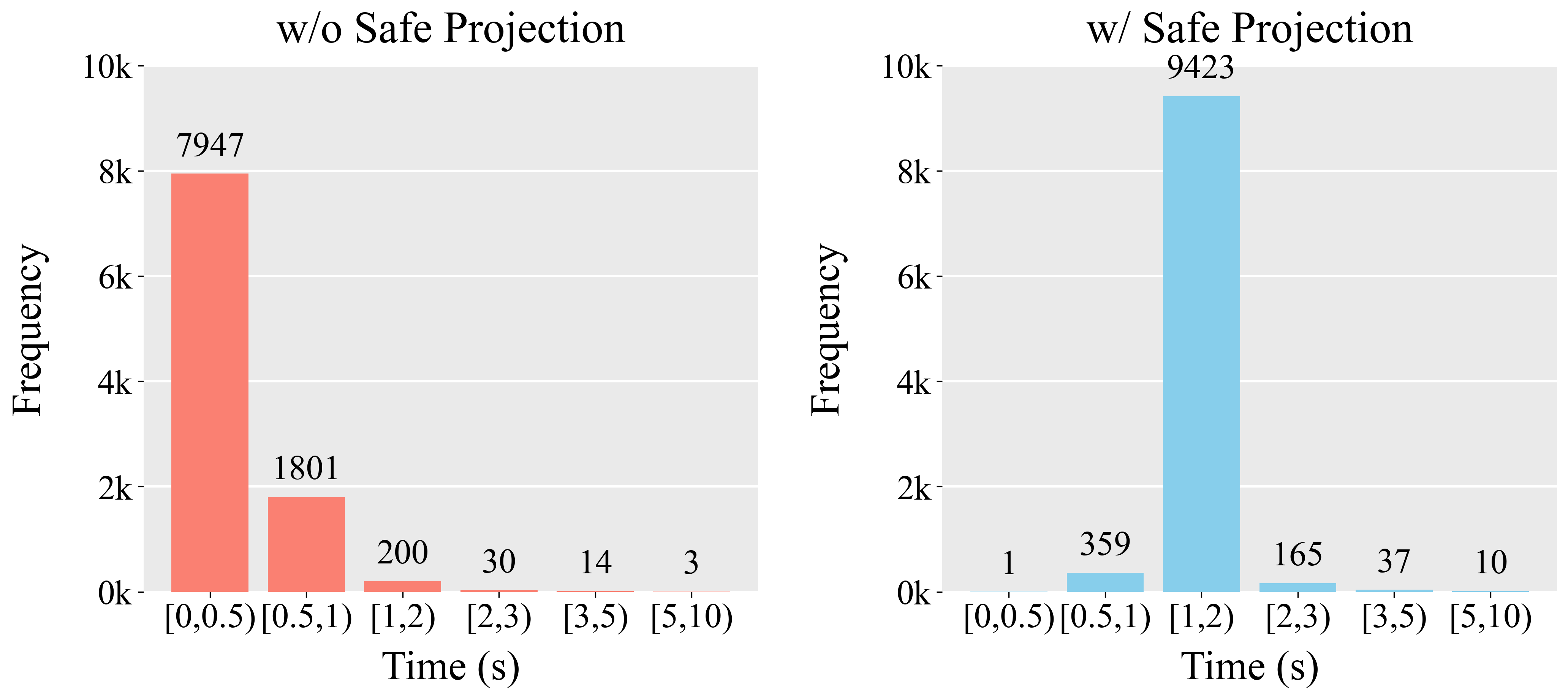}
\end{center}
  \vspace{-1.0em}
  \caption{Runtime distribution of our pipeline over the entire Thingi10K dataset. Note that we exclude 5 corrupted input meshes in the dataset.}
  \vspace{-2.0em}
\label{fig:runtime_10k}
\end{figure}

\paragraph{Execution time distribution} We evaluate the speed of our algorithm over the entire Thingi10K dataset and summarize the runtime distribution in Figure~\ref{fig:runtime_10k}. When reducing the number of triangles to 1$\%$, over 98\% meshes are processed in under 2 seconds. When not including the safe projection stage, over 97\% of meshes can be processed in under 1 second.


\subsection{Remeshing}

\paragraph{Remeshing performance}
We present a comparative analysis of our remeshing algorithm's performance against baseline methods in Table~\ref{tab:speed}. This analysis was conducted using a subset sampled from Thingi10k, comprising 124 non-watertight objects with self-intersections. On average, our remeshing algorithm processes a mesh with over 100k vertices in 3 ms.

\setlength{\tabcolsep}{4pt}
\begin{table}[htbp]
\small
  \centering
  \caption{Speed comparison between our remeshing algorithm and other methods with similar output resolutions. Reported numbers are averages over 124 non-watertight input meshes with self-intersections.}
    \begin{tabular}{l|cccccc}
    \toprule
    Method  & Manifold & ManifoldPlus & AlphaWrap. & OpenVDB & Ours \\
    \midrule
    \# V & 147k & 78k & 154k & 116k & 123k \\
    \# F & 294k & 155k & 309k & 232k & 245k \\
    Time (ms) & 896.69 & 3844.02 & 32843.28 & 62.40 & \textbf{2.97} \\
    \bottomrule
    \end{tabular}
  \label{tab:speed}
\end{table}

\paragraph{Self-intersection correction}\label{sec:inter_correct}
We test our revised DualMC remeshing algorithm on a large dataset to experimentally verify whether it guarantees intersection-free results. Specifically, we evaluate it on 10k randomly initialized $64^3$ grids, which are significantly more challenging than any SDF converted from objects. Among the 10k test cases, the original DualMC generates 8 cases with self-intersections, whereas our revised DualMC ensures intersection-free results across all cases.

\subsection{Mesh Simplification}
\label{sec:ablate_simp}

\begin{figure}[htbp]
\begin{center}
   \includegraphics[width=\linewidth]{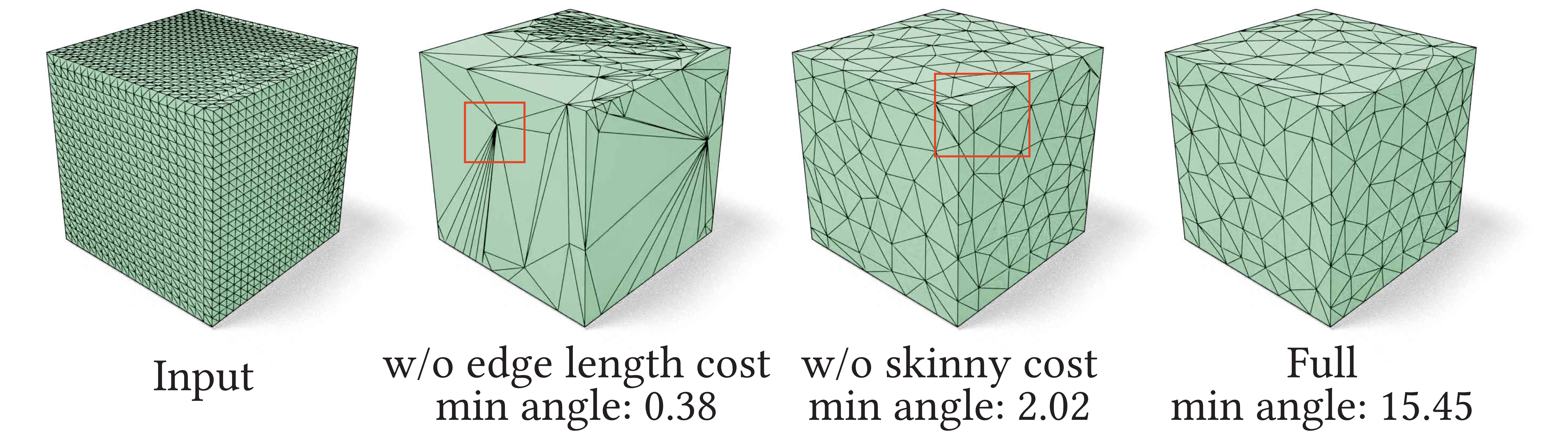}
\end{center}
\vspace{-0.5em}
  \caption{\textbf{Ablation study on the extra metrics in the edge computation.} Both edge length and skinny cost metrics can effectively reduce the appearance of skinny triangles. We assess their improvements by calculating the minimum angle of triangles on the output mesh.}
\label{fig:edge_cost}
\end{figure}

\paragraph{Effect of edge costs}
We demonstrate how including edge length cost $C_e$ and skinny cost $C_s$ impacts the simplification process. We evaluate the bad triangle ratio on the 100-mesh subset, where a triangle is defined as ``bad'' if its minimum angle is less than a pre-defined threshold $\alpha$. As shown in Table~\ref{tab:edge_cost}, including both costs can achieve the lowest triangle ratio in three threshold choices. Additionally, as shown in Figure~\ref{fig:edge_cost}, we find that these additional costs are essential in flat mesh regions where QEM score is constant, encouraging the generation of triangles with similar sizes and edge lengths.

\setlength{\tabcolsep}{4pt}
\begin{table}[htbp]
\small
  \centering
  \caption{Ablation study of edge cost in parallel mesh simplification. The bad triangle ratio is used as the metric. Note that a triangle with a minimum angle of less than the threshold angle $\alpha$ is regarded as a bad triangle. ($C_\mathrm{e}$ is the edge length cost, $C_\mathrm{s}$ is the skinny cost). }
    \begin{tabular}{l|cccc}
    \toprule
       & w/o $C_\mathrm{e}$ and $C_\mathrm{s}$  & w/o $C_\mathrm{e}$& w/o $C_\mathrm{s}$ & Full \\
    \midrule
     $\alpha=10^\circ$  & 13.16\%&13.32\% &3.72\% &\textbf{2.81\%} \\
     $\alpha=5^\circ$  & 4.89\%&4.84\% &0.17\% &\textbf{0.14\%} \\
     $\alpha=1^\circ$  & 0.56\%&0.53\% &0.04\% &\textbf{0.02\%} \\
    \bottomrule
    \end{tabular}
  \label{tab:edge_cost}
\end{table}

\begin{figure}[htbp]
  \centering
  \includegraphics[width=1.0\linewidth]{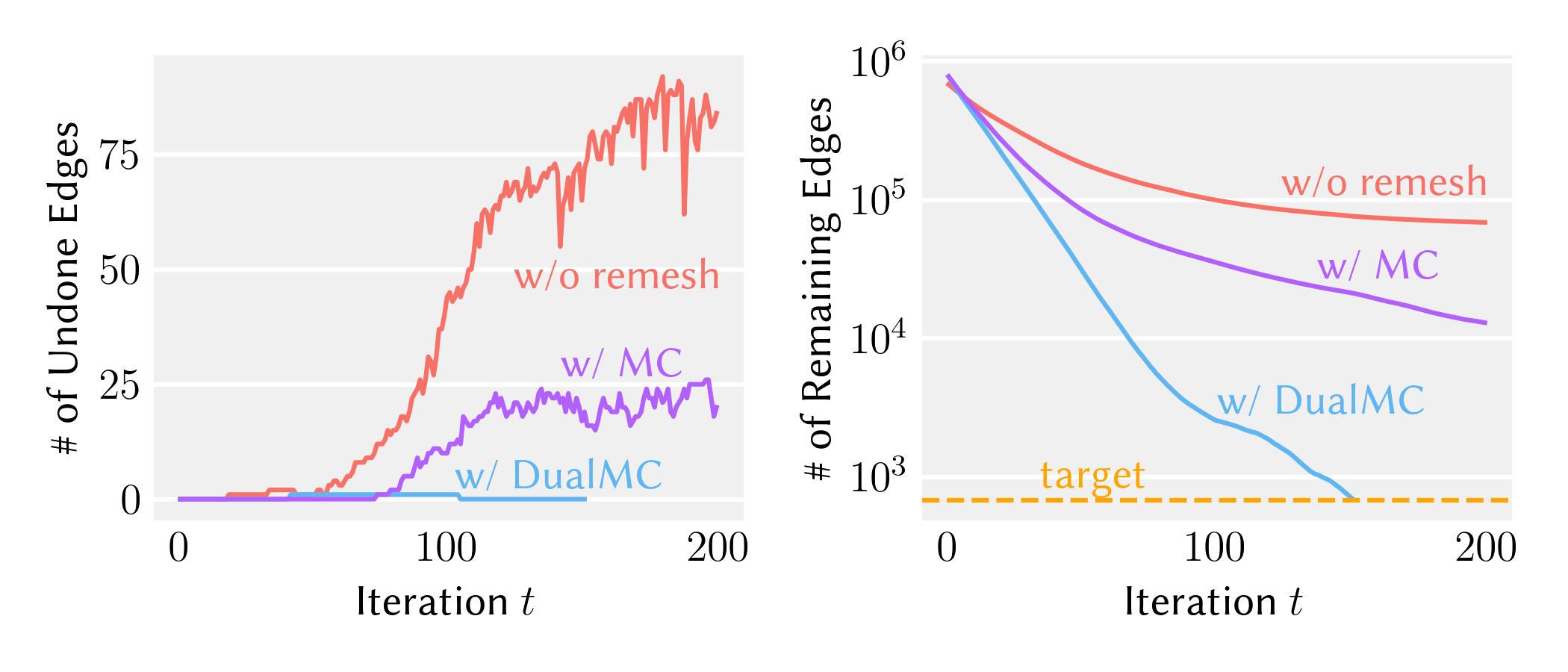}
  \vspace{-2.5em}
  \caption{\textbf{Impact of remeshing on simplification performance.} \textbf{Left:}
  number of edge collapses reversed by undo operation each iteration. \textbf{Right:} Number of edges remaining after each parallel edge collapse iteration. }
\label{fig:undo_remesh}
\end{figure}

\paragraph{Impact of remeshing on simplification}
We observe that our DualMC-based remeshing can improve the speed of our simplification stage by reducing the number of undo operations required. We conduct tests on a mesh with 450k faces, which includes many skinny triangles, to observe the differences in the simplification process with and without our remeshing. We also compare with MC-based remeshing, which replaces the DualMC in our remeshing by Marching Cubes, as shown in Figure~\ref{fig:undo_remesh}. We notice that even though the number of edges increases from 680k to 780k after remeshing, the triangle reduction speed is much faster than without remeshing or using MC-based remeshing. Particularly after 150 iterations, the process without remeshing or using MC-based remeshing stalls because many collapsing operations are undone due to the introduction of self-intersections.

\paragraph{Robustness of self-intersection check algorithm}
We evaluate the performance of our self-intersection check algorithm by comparing it with several prominent libraries, including libigl~\cite{jacobson2013libigl} and Blender~\cite{blender}.
We use CGAL~\cite{fabri2009cgal} to generate the ground-truth labels for the self-intersection check, as it is the most well-maintained 3D library. As demonstrated in Table~\ref{tab:selfintersect_runtime}, our algorithm is significantly more efficient and particularly robust to coplanar triangle intersections, where libigl and Blender tend to produce false negatives.
Moreover, we curated a dataset of 10,100 randomly generated triangle pairs, which includes coplanar triangles, sharp triangles, vertex-sharing cases, and edge-sharing cases, to evaluate different self-intersection algorithms. We present the quantitative results using this test set in Table~\ref{tab:selfintersect_quant}. We also compare our method with ``Ours (float)'', which indicates that all operations were performed using single precision instead of mixed precision.
By allowing a few false positive cases, our algorithm can achieve a 0$\%$ FNR, which is critical for guaranteeing self-intersection-free results, and we found that using floating point only cannot well handle all the intersection cases.


\begin{table}[htbp]
\small
  \centering
  \caption{Performance comparison between our triangle self-intersection checking algorithm with other libraries. Our algorithm handles vertex-sharing cases correctly and takes significantly less time to execute. Tested on three meshes, each with a different number of faces.}
    \begin{tabular}{lcccc}
    \toprule
        Method & \multicolumn{3}{c}{Time / ms} & Coplanar Detection\\
        \cline{2-4}
        & 2.5M & 1.1M & 250k & \\
    \midrule
    CGAL & 74631.3 & 42099.5&10186.4 & \cmark \\
    libigl & 7521.3& 3605.5& 828.3& \xmark \\
    Blender & 2770.1& 989.9& 255.0& \xmark \\
    \midrule
    Ours & \textbf{48.3} & \textbf{26.1} & \textbf{6.7} & \cmark \\
    \bottomrule
    \end{tabular}
  \label{tab:selfintersect_runtime}
\end{table}

\begin{table}[htbp]
\small
\centering
\caption{Quantitative comparison of self-intersecting triangle detection algorithms. We report the Accuracy, Recall, False Positive Rate (FPR) and False Negative Rate (FNR). Ours (float) indicates that all operations were performed using single precision.}
\label{table:comparison}
\begin{tabular}{@{}l|cccc@{}}
\toprule
Method & Accuracy$\uparrow$ & Recall$\uparrow$ &FPR$\downarrow$ & FNR$\downarrow$ \\
\midrule
libigl & \textbf{0.9991} &0.9950 & \textbf{0.0000}&0.0050 \\
Blender &0.9988   &0.9933 &\textbf{0.0000} &0.0067 \\
\midrule
Ours (float) &0.9903   &0.9983 &0.0114 &0.0017 \\
Ours &0.9906  &\textbf{1.0000} &0.0114 &\textbf{0.0000} \\
\bottomrule
\end{tabular}
\label{tab:selfintersect_quant}
\end{table}

\paragraph{Undo iterations}
To assess the frequency of undo operations needed to generate a self-intersection-free mesh during the simplification process, we recorded the number of undo operations performed in each edge-collapse iteration. We report the results on the 100-mesh subset and the full Thingi10K dataset. The results are summarized in Table~\ref{tab:worst}. In both tests, intersections introduced in about 99$\%$ of parallel edge collapses are resolved within one undo operation. Out of a total of 2.2 million edge-collapse iterations, only about 20 require three undo operations, and we have not encountered any case that requires four or more undo operations.

\setlength{\tabcolsep}{4pt}
\begin{table}[htbp]
\small
  \centering
  \caption{Number of parallel undo operations required after each parallel edge collapse on the \textit{entire} Thingi10K.}
    \begin{tabular}{l|cccccc}
    \toprule
      & 0 & 1 & 2 & 3 & more than 3 \\
    \midrule
    100 subset & 66.55\% & 33.45\% & 0.00\% & 0.00\% & 0.00\% \\
    10K dataset & 72.15\% & 26.62\% & 1.22\% & 0.0007\% & 0.00\% \\
    \bottomrule
    \end{tabular}
  \label{tab:worst}
\end{table}

\begin{figure}[htbp]
\centering
\includegraphics[width=\linewidth]{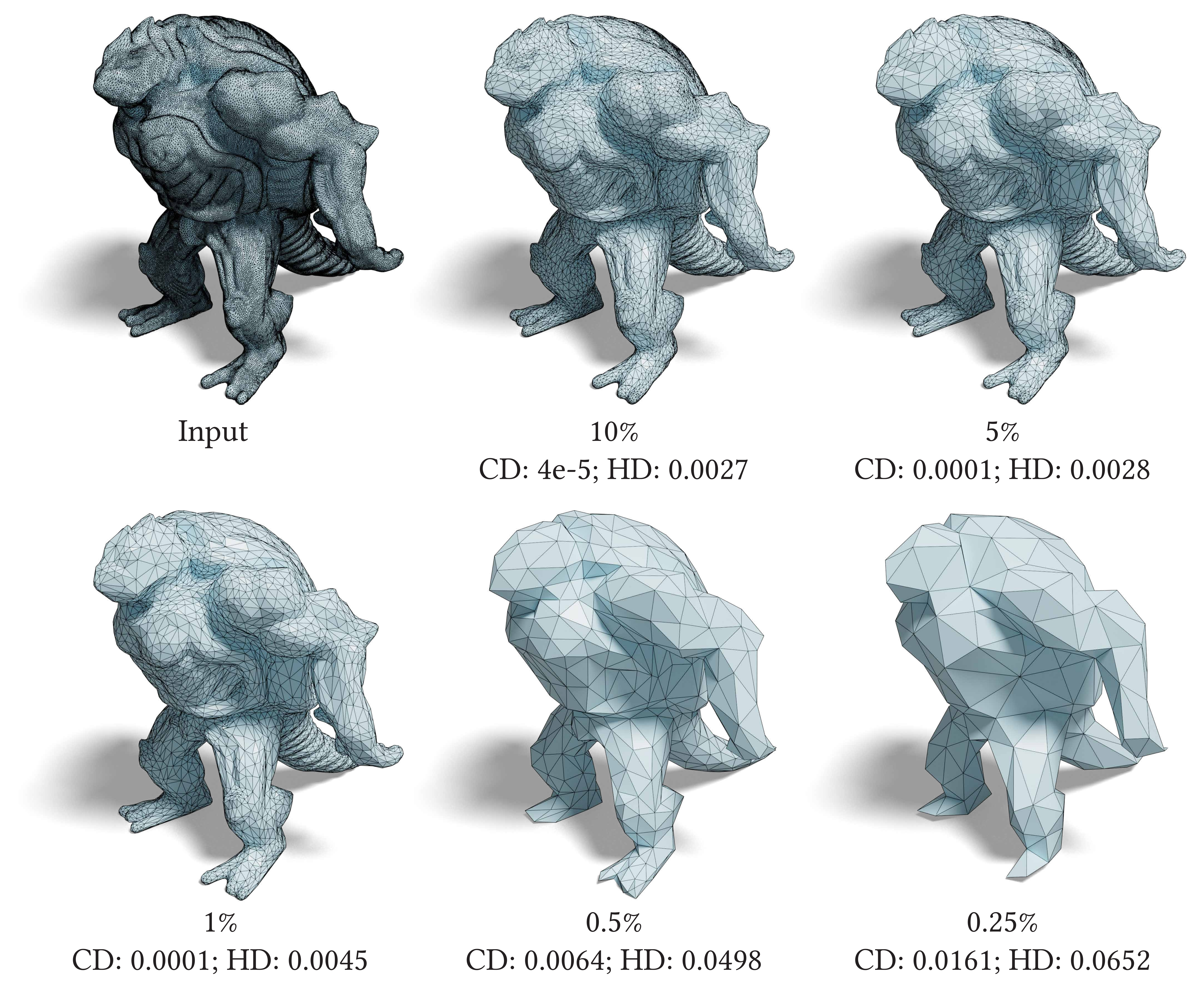}

\vspace{-0.5em}
  \caption{\textbf{Level of detail.} We examined the results of Chamfer Distance (CD) and Hausdorff Distance (HD) by applying different levels of detail to the same mesh.} 
  \label{fig:threshold_comp}
  \vspace{-1em}
\end{figure}

\paragraph{Level of detail}
We evaluate our method's performance with varying decimation ratios, visualized in Figure~\ref{fig:threshold_comp}. Notably, when we reduce the mesh to below 1\% of its original face count, there is a significant increase in CD and HD metrics due to more frequent self-intersections and invalid collapses in smaller meshes. As outlined in Section 5.3, to address these issues, we accumulate problematic edges in an invalid edge list for a certain number of iterations until reaching a specified tolerance. This list is crucial for avoiding the selection of edges that would compromise mesh integrity during further decimation. By focusing on edges that maintain valid configurations, we effectively control quality at extreme decimation levels, although this approach does come with a slightly higher computational cost. This selective strategy explains the observed increase in distance metrics as the decimation rate is less than 1\%.


\subsection{Safe Projection} \label{sec:analysis_stage3}

\begin{figure}[htbp]
\begin{center}
   \includegraphics[width=\linewidth]{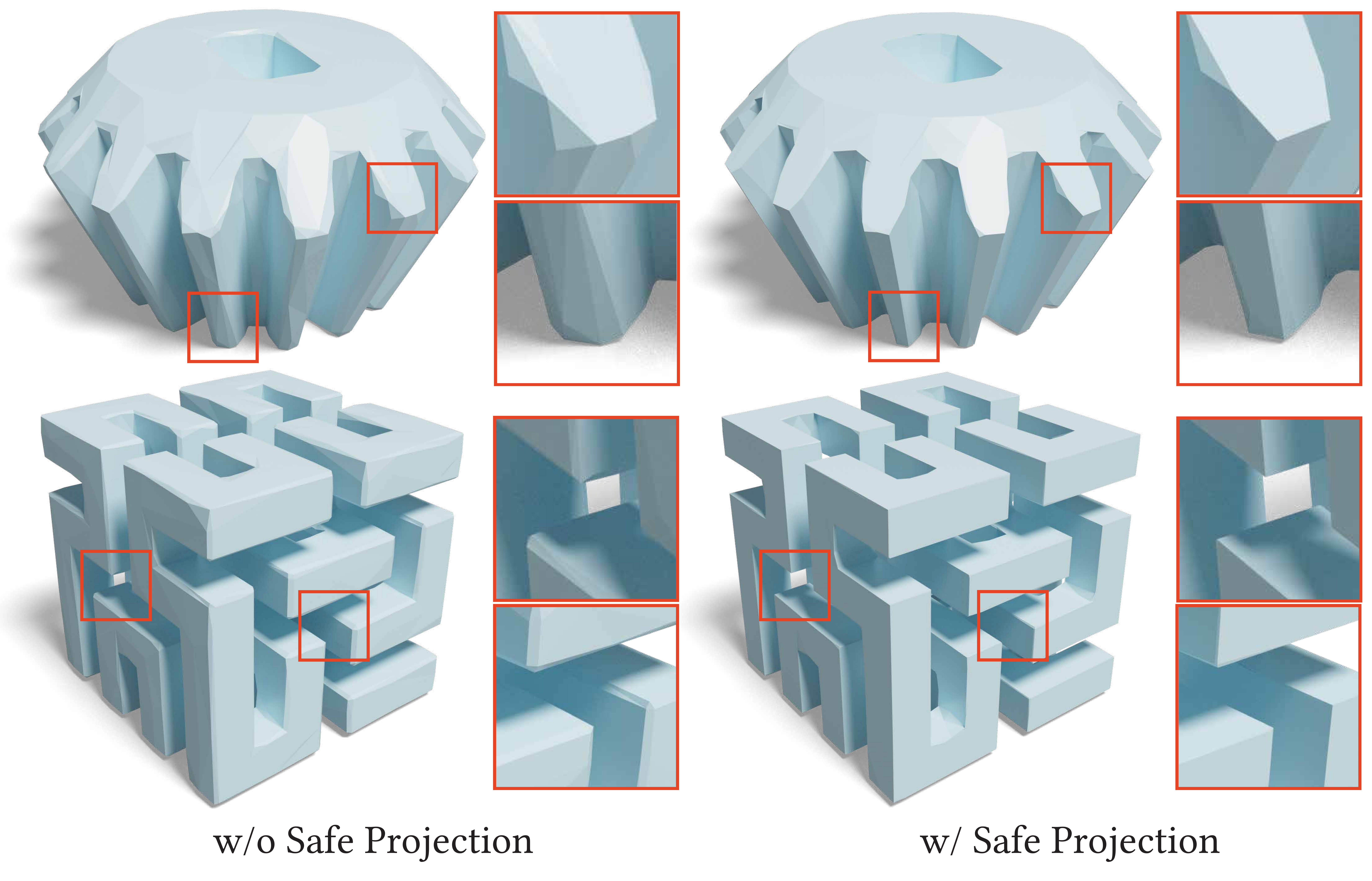}
\end{center}
\vspace{-0.5em}
\caption{\textbf{Effectiveness of safe projection. }Safe projection recovers sharp corners and edges from our simplified mesh, matching input geometry more closely.}
\vspace{-0.6em}
  \label{fig:sharp_feat}
\end{figure}

\paragraph{Effectiveness of safe projection} We have quantitatively shown in Table~\ref{tab:baseline_comp} that the safe projection (stage 3) of our method improves alignment with the input mesh. We further demonstrate in Figure~\ref{fig:sharp_feat} that the safe projection stage recovers sharp corners and edges from our simplified mesh.

\begin{figure}[htbp]
\begin{center}
\vspace{-1em}
   \includegraphics[width=0.7\linewidth]{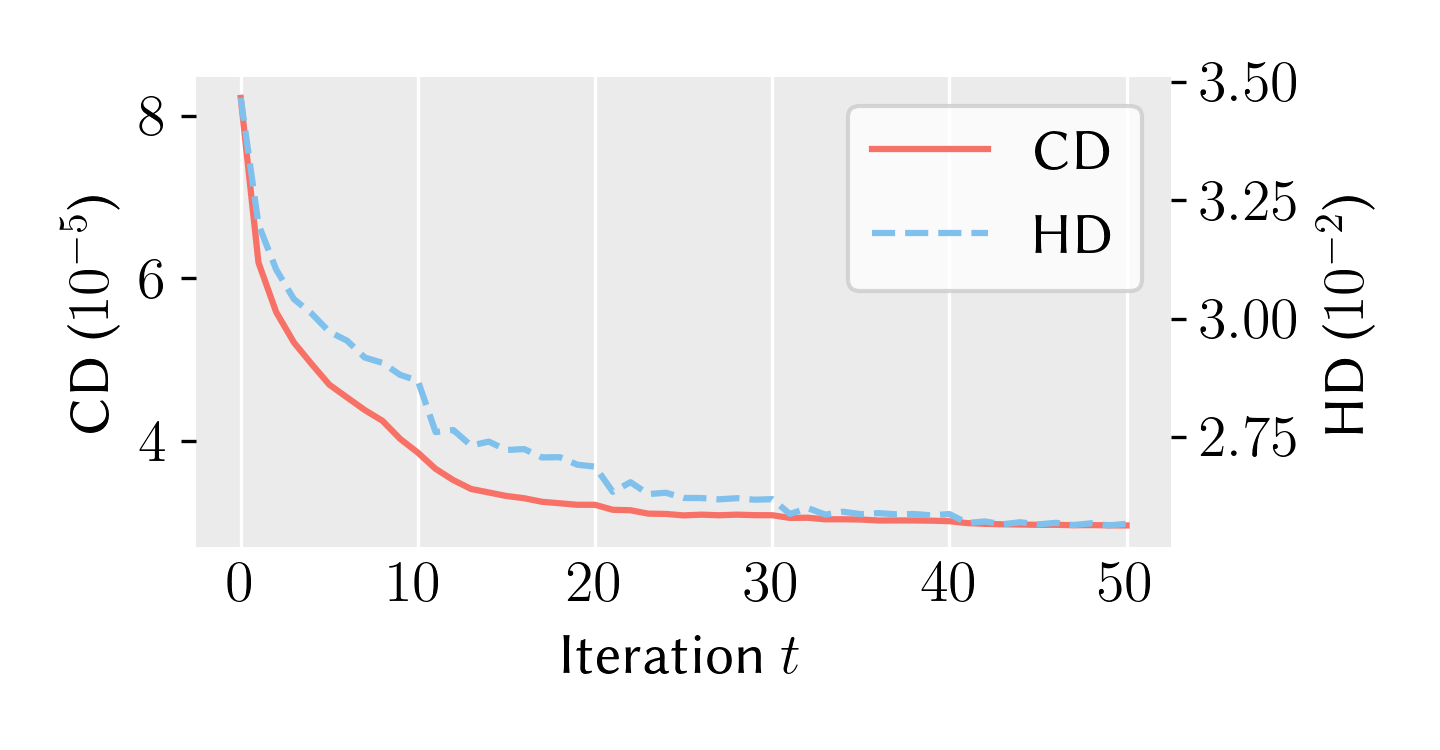}
\end{center}
  \vspace{-1.5em}
  \caption{Change of distances during optimization.}
  \label{fig:stage3_convergence}
  \vspace{-0.5em}
\end{figure}

\begin{figure*}[htbp]
\begin{center}
   \includegraphics[width=\textwidth]{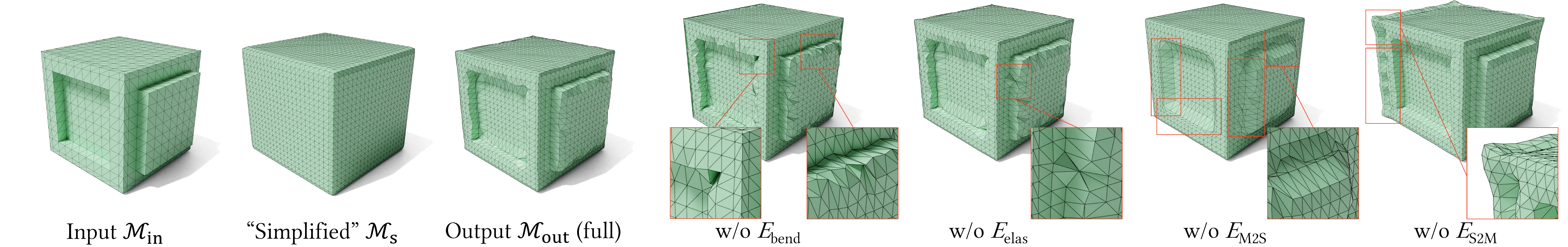}
\end{center}
  \caption{\textbf{Ablation study on the energies in safe projection}. The absence of specific energies results in distinct artifacts in the output: absence of $E_\mathrm{bend}$ leads to large bending angles; absence of $E_\mathrm{elas}$ produces more skinny triangles; without $E_\mathrm{M2S}$, there is insufficient recovery of ground truth details; and without $E_\mathrm{S2M}$, parts of the simplified mesh are omitted during the projection. With all four energies, ours outperformed all others but failed to fully recover the sharp features due to the extreme loss of details in the hypothetical simplified mesh, $\mathcal M_\mathrm{s}$.
  }
  \label{fig:ablation_energy}
\end{figure*}

\setlength{\tabcolsep}{4pt}
\begin{table}[htbp]
\small
  \centering
  \caption{
      Ablation study on the energies in safe projection. Tested on the output of our simplification stage on the same 100-mesh subset of Thingi10K as Table \ref{tab:baseline_comp}. 
  }
    \begin{tabular}{l|cccccc}
    \toprule
     & w/o $E_\mathrm{bend}$ & w/o $E_\mathrm{elas}$ & w/o $E_\mathrm{M2S}$ & w/o $E_\mathrm{S2M}$ & Ours \\
    \midrule
    CD $\downarrow$ ($10^{-5}$) & 3.149          & 4.651 & 6.157 & 6.093 & \textbf{2.899} \\
    HD $\downarrow$ ($10^{-2}$) & 2.562          & 2.952 & 3.274 & 3.134 & \textbf{2.495} \\
    $M_t$ $\uparrow$ (deg)    & \textbf{3.186} & 2.546 & 3.026 & 2.858 & 2.647 \\
    \bottomrule
    \end{tabular}
  \label{tab:stage3_ablation}
\end{table}

\paragraph{Energy functions} Table~\ref{tab:stage3_ablation} presents the results of our ablation study on the energy functions proposed in Section~\ref{sec:energy}. We observe that both $E_\mathrm{M2S}$ and $E_\mathrm{S2M}$ are crucial for reducing CD and HD. Furthermore, $E_\mathrm{bend}$ and $E_\mathrm{elas}$ not only enhance the shape quality of the output mesh but also slightly reduce the CD and HD. We believe that $E_\mathrm{bend}$ and $E_\mathrm{elas}$ can help avoid those local minima associated with large local deformations of the simplified mesh. To better illustrate the impact of each energy function, we designed a hypothetical scenario using a pair of ``input mesh'' $\mathcal M_\mathrm{in}$ and ``simplified mesh'' $\mathcal M_\mathrm{s}$. We applied our safe projection stage to this fictitious pair to align the input with the ground truth and visualized the outcomes when using different energies in Figure~\ref{fig:ablation_energy}. While the actual simplified meshes generated in our simplification stage (stage 2) generally resemble the ground truth closely, we intentionally made the fictitious input very different from the ground truth to increase the challenge of shape recovery and highlight potential artifacts. Figure~\ref{fig:ablation_energy} shows different artifacts that emerge when each energy function is omitted.

\paragraph{Convergence of the Newton-type solver} We visualize the change of CD and HD during the optimization in Figure~\ref{fig:stage3_convergence}. The distances are averaged over the 100-mesh subset. We conclude that the CD and HD distances are consistently reduced while we optimize the augmented energy $B(X)$, and 30--50 iterations are sufficient for our Newton-type solver to converge under these conditions.


\section{Limitations and Future work}
Our method excels in handling complex meshes with a large number of faces but does not offer a significant speed advantage when simplifying smaller meshes. This discrepancy arises because edge collapsing often leads to self-intersections in smaller meshes that are not remeshed, whereas remeshing small meshes significantly increases the face count, which compromises efficiency compared to directly simplifying smaller meshes with other methods.

Our method, particularly in Stage 1, does not assume any manifold properties of the input mesh and performs remeshing on all inputs to produce manifold meshes. However, when the input mesh is already manifold and exhibits certain topological features, the remeshing stage may inadvertently remove these characteristics—for example, tiny holes or gaps smaller than the DualMC voxels. For cloth meshes, our method may generate two overlapping layers. A potential extension is to use a normal-based face selection strategy to remove the extra layer, thereby preserving a single-layer mesh that closely matches the original shape while maintaining high mesh quality.

In our current pipeline, the simplification of mesh connectivity (stage 2) and the modification of mesh vertex positions (stage 3) are fully decoupled. Stage 2, which solely takes $\mathcal M_\mathrm{r}$ as input, is agnostic of the input shape $\mathcal M_\mathrm{in}$ and therefore cannot optimize $\mathcal M_\mathrm{s}$ to reduce its distance from $\mathcal M_\mathrm{in}$. Conversely, while stage 3 aligns its input $\mathcal M_\mathrm{s}$ with $\mathcal M_\mathrm{in}$, the mesh connectivity can no longer change in this stage, which restricts the ability of stage 3 to restore sharp features. A potential avenue for future research is to integrate mesh simplification with safe projection, enabling dynamic optimization of both connectivity and vertex positions while ensuring the mesh remains manifold, watertight, and free of self-intersections. 





\section{Conclusion}

In this paper, we introduced PaMO, a novel GPU-based mesh optimization algorithm capable of efficiently converting arbitrary input meshes into low-poly, manifold, intersection-free forms suitable for real-time applications in graphics pipelines. Through a detailed exposition of our three-stage process---remeshing, simplification, and safe projection---we have demonstrated that our method not only generates high-quality meshes but also significantly enhances the processing speed compared to existing methods.

Our evaluations using the Thingi10K dataset have consistently shown that PaMO can produce superior results in terms of geometric properties and mesh fidelity. The ability of our algorithm to maintain manifold properties and eliminate self-intersections while operating under the computational limits of modern GPU architectures is particularly noteworthy. This capability makes PaMO an attractive solution for high-efficiency requirements in industries such as gaming, AR/VR, and digital fabrication.

By pushing the boundaries of what is possible in mesh optimization, this work contributes to the foundational technology necessary for the next generation of real-time 3D content creation and manipulation, paving the way for more immersive and dynamic digital environments.

\section*{Acknowledgement}
This research was funded by Hillbot Inc. Hao Su is the CTO for Hillbot and receives income. The terms of this arrangement have been reviewed and approved by the University of California, San Diego in accordance with its conflict of interest policies.

\bibliographystyle{eg-alpha-doi}  
\bibliography{egbibsample} 

\clearpage

\section*{Supplementary Materials}
\section{Intersection-free Dual Marching Cubes}

\subsection{Look-Up Table of DualMC}

\begin{figure}[ht]
\centering
   \includegraphics[width=\linewidth]{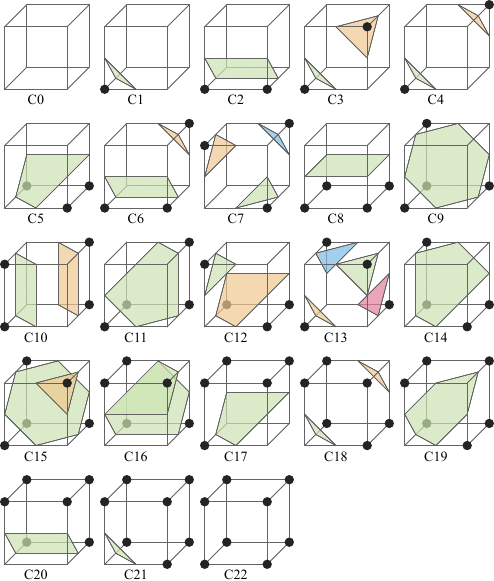}
  \caption{Look-Up table of DualMC cases. Figure adapted from \protect\citesupp{nielson2004dual}.}
\label{fig:mctable}
\end{figure}

Here we show the look-up table for building DualMC patches. The original DualMC~\citesupp{nielson2004dual} may introduce non-manifold structures in very rare cases, we follow the solution in ~\citesupp{wenger2013isosurfaces} to solve the problematic C16 and C19 cases in which ambiguous faces are shared.

\section{Detailed Algorithms}

\subsection{DualMC Self-intersection Correction}

\begin{figure}[ht]
\centering
   \includegraphics[width=0.6\linewidth]{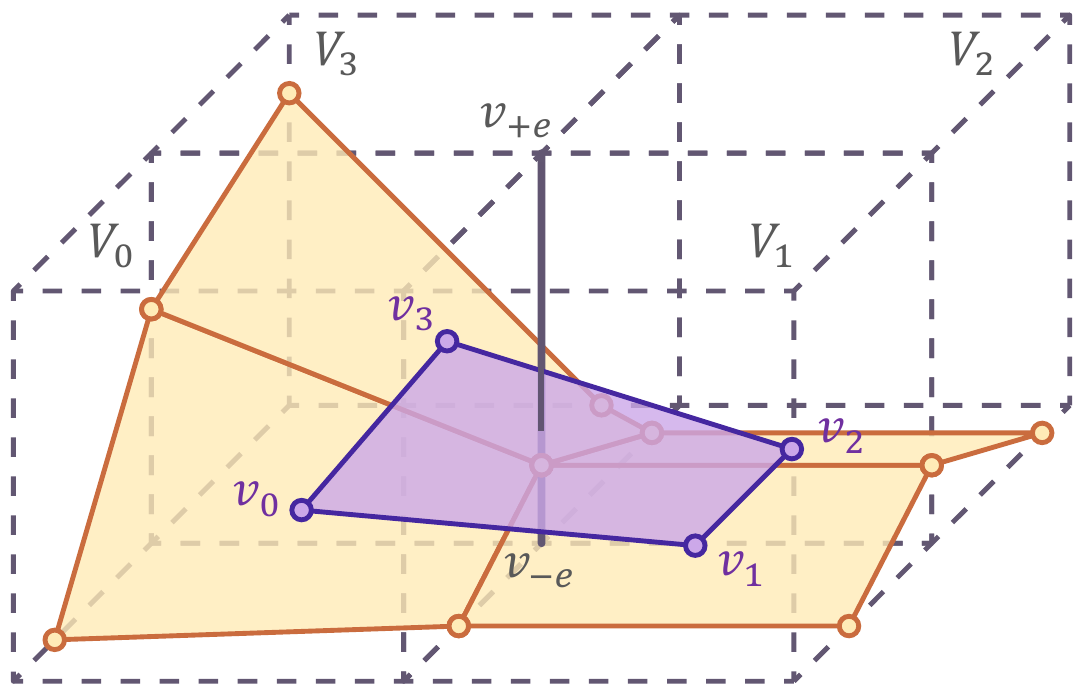}
\vspace{-1em}
  \caption{Patches (in yellow) and a quad (in purple) in DualMC. $(v_{+e}, v_{-e})$ is a valid edge. }
\label{fig:supp_dualmc}
\end{figure}

\paragraph{Handling numerical instability}
In DualMC, the patch vertex $v$ is determined through linear interpolation between the corresponding edge vertices ($v_e^0$, $v_e^1$) based on their SDF values:

\begin{align}
v = v_e^0 + t (v_e^1 - v_e^0), \quad t &= \frac{- f(v_e^0)}{f(v_e^1) - f(v_e^0)}, \label{eq:t}
\end{align}
where $ f(\cdot) $ represents the SDF function and $t$ is within $[0, 1]$. 

We found extreme $t$ values can introduce numerical instability and cause inter-quad intersections. To address this, we apply a sigmoid smoothing function to $ t $ and use $t'$ in the linear interpolation to mitigate extreme values:
\begin{align}
t' &= \frac{1}{1 + \exp\left(-\beta (t - 0.5)\right)} \label{eq:t_sig}
\end{align}
In this work, $ \beta = 5 $ is chosen based on experimental observations.

\paragraph{Quad Division}
We adopt the concept of the \textit{envelope} from~\citesupp{ju2006intersection,wang2009intersection} to eliminate intersections caused by quad division. As shown in Figure~\ref{fig:supp_dualmc}, a quad $ Q = (v_0, v_1, v_2, v_3) $ is defined around a valid edge $ e = (v_{+e}, v_{-e}) $. The envelope $ E_e $ is defined using $ v_{+e}, v_{-e}, v_0, v_1, v_2, v_3 $. The concavity of the vertices in $ Q $ is determined using the scalar triple product. A vertex $v_i$ is defined as a \textit{concave vertex} if either of the following condition holds:
\begin{align}
(v_i - v_{+e}) \cdot \left[ (v_i^l - v_{+e}) \times (v_i^r - v_{+e}) \right] < 0, \label{eq:concavity1} \\
(v_i - v_{-e}) \cdot \left[ (v_i^l - v_{-e}) \times (v_i^r - v_{-e}) \right] > 0. \label{eq:concavity2}
\end{align}
where $v_i^l$ and $v_i^r$ are the two adjacent vertices of $v_i$ in the quad.

The quad is divided based on the distribution of concave vertices:
\begin{itemize}
    \item If only one diagonal of $ Q $ contains concave vertices, the quad is divided along that diagonal.
    \item If both diagonals contain concave vertices, $ Q $ is divided into four triangles, with an additional vertex $ v_e $ placed on edge $ e $.
    \item If no concave vertices exist, we opt for a division that maximizes the minimal angle of each triangle.
\end{itemize}
When computing the additional $v_e$ in the second case, we determine its position using linear interpolation based on the edge vertices SDF values, combined with the sigmoid smoothing function to more accurately represent the iso-surface.

We implemented the quad division strategy using a 2-kernel Gather approach to ensure parallelism. First, we determine whether each vertex is concave and record the number of triangles to be generated within each quad along with the count of newly created vertices, and then perform the appropriate quad division accordingly.

\subsection{Computing the Energy Functions}

In the safe projection stage (stage 3), as we optimize the augmented energy function $B(X)$ using Newton's method, we need to compute the gradient $\nabla B(X)$ and the Hessian matrix $\nabla^2 B(X)$ at each iteration. In this section, we provide details about the computation of each term in $B(X)$ and their gradients and Hessian matrices. 

\paragraph{Distance energy $E_\mathrm{S2M}$} 
At the $t$-th iteration, we update the nearest neighboring point $\tilde y_i \in \mathcal M_\mathrm{in}$ for every $i \in V(\mathcal S)$: 
\begin{align}
    \tilde y_i(X^t) = \underset{y \in \mathcal M_\mathrm{in}}{\operatorname{argmin ~}} \|x^t_i - y\|^2.
\end{align}
With the nearest neighbors fixed, 
\begin{align}
    \tilde E_\mathrm{S2M}(X; \{\tilde y_i\}_{i=1}^n) = \sum_{i \in V(\mathcal S)} s^0_i \|x_i - \tilde y_i\|^2
\end{align}
in the following iterations as an estimation of $E_\mathrm{S2M}$ before we update the nearest neighbors again. The Hessian matrix is block-diagonal, where every block is $3\times 3$. 

\paragraph{Distance energy $E_\mathrm{M2S}$} 
At the $t$-th iteration, we update the nearest neighboring triangle $(\tilde i_s, \tilde j_s, \tilde k_s) \in F(\mathcal S)$ for every sample $y_s (1 \le s \le m)$:
\begin{align}
    (\tilde i_s, \tilde j_s, \tilde k_s)(X^t) = \underset{(i, j, k) \in F(\mathcal S)}{\operatorname{argmin ~}} d(\triangle_{ijk}(X^t), y_s) 
\end{align}
where $d(\triangle_{ijk}(X^t), y_s)$ is the distance between point $y_s$ and the triangle whose vertices are $(x^t_i, x^t_j, x^t_k)$. These point-triangle distances can further be classified into point-point distances, point-line distances, and point-plane distances according to the position of the nearest neighboring point of $y_s$ on $\triangle_{ijk}(X^t)$. Each distance category corresponds to a differentiable closed-form formula. We refer the reader to {Li2020IPC} for the detailed formulas. 
Similar to $E_\mathrm{S2M}$, we estimate $E_{M2S}$ as 
\begin{align}
    \tilde E_\mathrm{M2S}(X; \{(\tilde i_s, \tilde j_s, \tilde k_s)\}_{s=1}^m) = \frac{A(\mathcal M_\mathrm{in})}{m} \sum_{s=1}^{m} d(\triangle_{\tilde i_s\tilde j_s\tilde k_s}(X), y_s). 
\end{align}
The Hessian matrix $\nabla^2_X \tilde E_\mathrm{M2S}(X)$ is a block-sparse matrix, where every block is at most $9 \times 9$ and corresponds to a triangle $(\tilde i_s, \tilde j_s, \tilde k_s)$. 

\paragraph{Elastic energy $E_\mathrm{elas}$}
We refer the reader to \citesupp{10.1145/2343483.2343501} for the computation of the gradient and Hessian of the St. Venant-Kirchhoff energy. The Hessian matrix $\nabla^2_X E_\mathrm{elas}(X)$ is a block-sparse matrix, where every block is at most $9 \times 9$ and corresponds to a triangle $(\tilde i_s, \tilde j_s, \tilde k_s)$. 

\paragraph{Bending energy $E_\mathrm{bend}$}
We refer the reader to \citesupp{tamstorf_discrete_2013} for the computation of the gradient and Hessian of the bending energy. The Hessian matrix $\nabla^2_X E_\mathrm{bend}(X)$ is a block-sparse matrix, where every block is at most $12 \times 12$ and involves 4 vertices of $\mathcal S$ belonging to two adjacent triangles. 

\paragraph{Barrier function $b(d_c(X))$}
We refer the reader to \citesupp{Li2020IPC} for the computation of the gradient and Hessian of the barrier function. The Hessian matrix $\nabla^2_X \sum_{c \in C} b(d_c(X))$ is a block-sparse matrix, where every block is at most $12 \times 12$ and involves 4 vertices of $\mathcal S$. The 4 vertices correspond to an activated ($d_c < \hat d$) contact pair, which is either a point-triangle pair or an edge-edge pair.

\section{Intersection tests}
\subsection{Coplanar triangle intersection analysis}

Coplanar triangle intersections in libigl and Blender often struggle to differentiate between non-intersecting adjacent pairs that share vertices and actual self-intersections. To address this issue, we have implemented a 2D intersection test that categorizes triangle pairs based on the number of shared vertices. The complete algorithm for our self-intersection detection is shown in Algorithm~\ref{alg:cuselfinsec}. Firstly, we construct a BVH tree to accelerate our triangle-triangle query process (Line~\ref{alg:bvh}). For each triangle in the mesh, we compute its intersection candidates by querying the BVH tree and checking for AABB overlaps (Line~\ref{alg:candidate}). Then, for non-coplanar cases, we utilize \textit{Devillers' algorithm} to detect intersections; for coplanar cases, we divide the algorithm into three scenarios (Line~\ref{alg:coplanar_start} - Line~\ref{alg:coplanar_end}) based on the number of shared vertices.

\begin{algorithm}
\caption{Parallel Self-Intersection Detection \\
$F(\mathcal M)$ denotes the face set.}
\label{alg:cuselfinsec}
\DontPrintSemicolon
\SetKwComment{Comment}{/* }{ */}
\SetKwProg{Fn}{Procedure}{:}{}
\SetKwFunction{FCPI}{CoplanarIntersect}
\Fn(\tcp*[h]{App. Sec. 3.1}){\FCPI{$T, C, s$}}{
     \uIf(\tcp*[h]{Fig. 7(1)}){$s = 0$}{
        \label{alg:coplanar_start}
        \KwRet $\operatorname{CoplanarNonShareIntersect}(T, C)$  
     }
     \uElseIf(\tcp*[h]{Fig. 7(2)(3)}){$s = 1$} {
        \KwRet $\operatorname{AngleOverlapIntersect}(T, C)$  
     }
     \ElseIf(\tcp*[h]{Fig. 7(4)(5)}){$s = 2$} {
        \KwRet $\operatorname{SameSideIntersect}(T, C)$ 
        \label{alg:coplanar_end}
     }
     \KwRet True
}
\SetKwFunction{FMain}{IntersectionCheck}
\Fn{\FMain{$\mathcal M$}}{
    \KwInput{Mesh $\mathcal M$}
    \KwOutput{Self-intersecting triangle pairs $\mathcal{T}$}
    $\mathrm{BVH} \gets \operatorname{ConstructLBVHTree}(\mathcal M)$ \tcp*[h]{parallel}\;
    \label{alg:bvh}
    \ForEach(\tcp*[h]{parallel}){$T \in F(\mathcal M)$}{
        $\mathcal C_T \gets \operatorname{BVHQuery}(\mathrm{BVH}, T)$ \tcp*[h]{intersection candidates}
        \label{alg:candidate}
    }
    \ForEach(\tcp*[h]{parallel}){$(T, C)$ s.t. $T \in F(\mathcal M), C \in \mathcal C_T$}{
            $s \gets \operatorname{CountSharedVertices}(T, C)$\;
            \uIf{$\operatorname{IsCoplanar}(T, C)$\label{alg:is_coplanar} } 
            {
                \If{\FCPI{$T, C, s$}}{
                    $\mathcal{T} \gets \mathcal{T} \cup \{(T, C)\}$
                }
            } \ElseIf{$s \ne 2$}{
                $l \gets \operatorname{GetIntersectLine}(T, C)$\; \label{alg:get_intersect_line} 
                \If(\tcp*[h]{App. Sec. 3.2}){$\|l\|>0 \land \neg (\|l\| \leq \epsilon \land s = 1)$}{
                    $\mathcal{T} \gets \mathcal{T} \cup \{(T, C)\}$
                }
            }
    }
    \KwRet $\mathcal{T}$
}
\end{algorithm}



Here we explain the details of how we handle the detection of intersections among coplanar triangles.

\paragraph{Coplanar with no shared vertex}
We also use \textit{Devillers' algorithm} to detect intersections; for those with one shared vertex

\paragraph{Single vertex sharing case}



\begin{wrapfigure}{l}{0.18\textwidth}
    \centering
    \vspace{-2em}
    \includegraphics[width=0.22\textwidth]{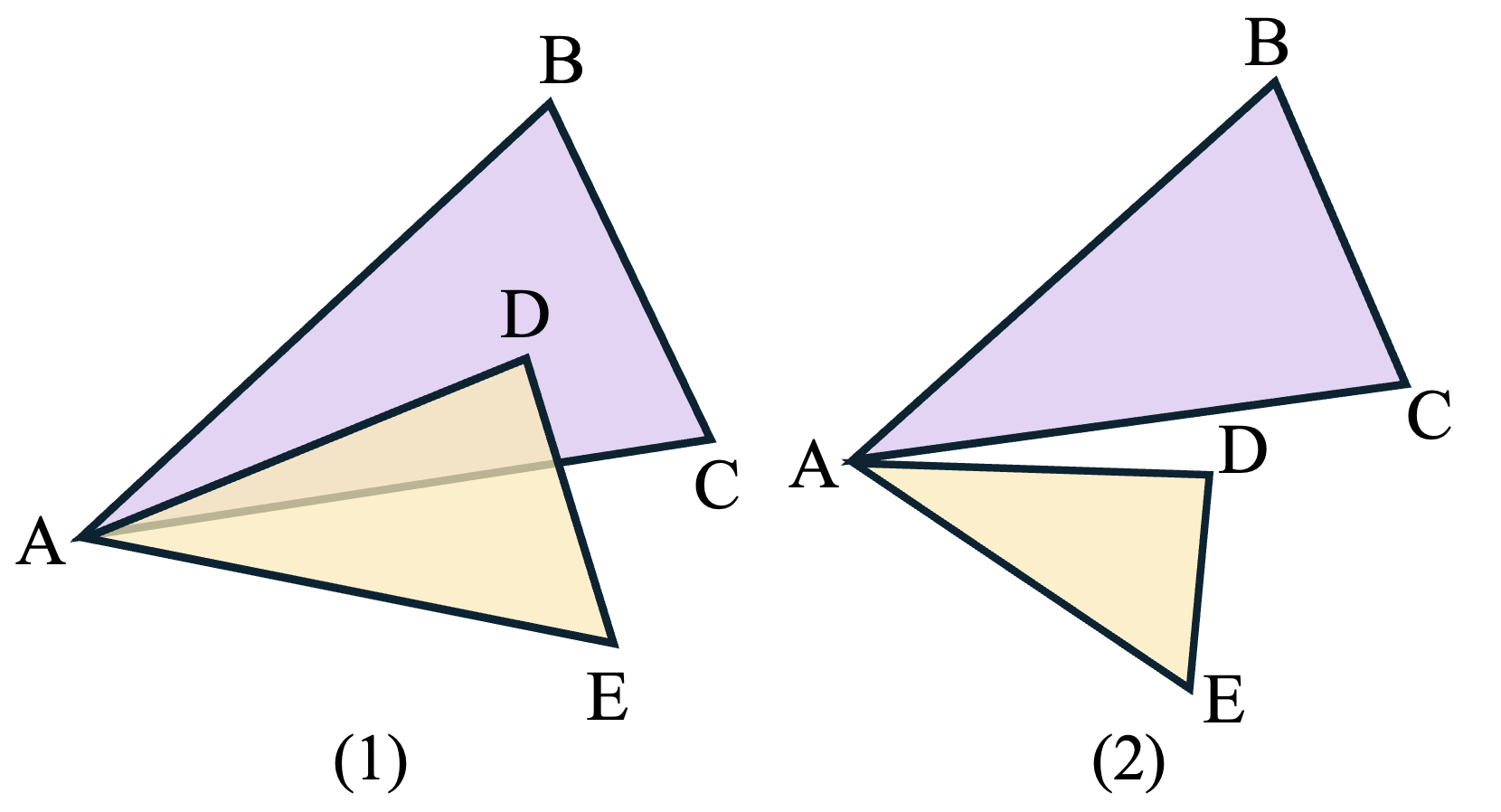}
    \vspace{-2em}
    \caption{Single-vertex-sharing cases.}
    \label{fig:single_vertex}
    \vspace{-2em}
\end{wrapfigure}

When there is one shared vertex, we propose \textit{angle-overlap test}. We begin by designating a shared vertex as \( A \) from triangles \( \triangle ABC \) and \( \triangle ADE \), as shown in Figure~\ref{fig:single_vertex}. We aim to determine whether an edge of one triangle, such as vector \( \overrightarrow{AD} \) or \( \overrightarrow{AE} \), lies between the edges \( \overrightarrow{AB} \) and \( \overrightarrow{AC} \) of the other triangle. 

Therefore, we first calculate the cross products for vectors \( \overrightarrow{AB} \), \( \overrightarrow{AC} \), \( \overrightarrow{AD} \), and \( \overrightarrow{AE} \) originating from \( A \). We store the results of these calculations in variables corresponding to each pair's cross product—specifically:
\begin{align*}
&\vec c_1 = \overrightarrow{AB} \times \overrightarrow{AD}, 
&\vec c_2 = \overrightarrow{AB} \times \overrightarrow{AE}, \\
&\vec c_3 = \overrightarrow{AC} \times \overrightarrow{AD}, 
&\vec c_4 = \overrightarrow{AC} \times \overrightarrow{AE}.
\end{align*}

By comparing the signs of \( \vec c_1 \), \( \vec c_2 \), \( \vec c_3 \), and \( \vec c_4 \), we can infer the following:
\begin{itemize}
    \item If \( \vec c_1 \cdot \vec c_3 \leq 0\), \( \overrightarrow{AD} \) may lie between \( \overrightarrow{AB} \) and \( \overrightarrow{AC} \).
    \item If \( \vec c_2 \cdot \vec c_4 \leq 0\), \( \overrightarrow{AE} \) may lie between \( \overrightarrow{AB} \) and \( \overrightarrow{AC} \).
    \item If \( \vec c_1 \cdot \vec c_2 \leq 0\), \( \overrightarrow{AB} \) may lie between \( \overrightarrow{AD} \) and \( \overrightarrow{AE} \).
    \item If \( \vec c_3 \cdot \vec c_4 \leq 0\), \( \overrightarrow{AC} \) may lie between \( \overrightarrow{AD} \) and \( \overrightarrow{AE} \).
\end{itemize}

While direction comparison can filter out most cases, the nature of the cross-product means that there can be instances where the directions differ but no actual overlap exists. To address this, after initial filtering using the directions of the $\vec c_i$, we further verify the situation by calculating the sum of the angles between the vector in question and the other two vectors. Specifically, if the directions differ, we compute this sum. If the sum of these angles is less than or equal to \( 180^\circ \), we confirm the presence of an intersection.

Let \( \theta_{AB, AD} \) be the angle between \( \overrightarrow{AB} \) and \( \overrightarrow{AD} \), 
\( \theta_{AB, AE} \) be the angle between \( \overrightarrow{AB} \) and \( \overrightarrow{AE} \), 
\( \theta_{AC, AD} \) be the angle between \( \overrightarrow{AC} \) and \( \overrightarrow{AD} \), 
and \( \theta_{AC, AE} \) be the angle between \( \overrightarrow{AC} \) and \( \overrightarrow{AE} \), where 
\begin{align*}
    \theta_{AB, AD}, \theta_{AB, AE}, \theta_{AC, AD}, \theta_{AC, AE} \in [0, \pi].
\end{align*}
The conditions for confirming overlap are given by:
\[
\begin{cases}
\theta_{AB, AD} + \theta_{AC, AD} < \pi, & \text{if } \vec c_1 \cdot \vec c_3 < 0, \\
\theta_{AB, AE} + \theta_{AC, AE} < \pi, & \text{if } \vec c_2 \cdot \vec c_4 < 0, \\
\theta_{AD, AB} + \theta_{AE, AB} < \pi, & \text{if } \vec c_1 \cdot \vec c_2 < 0, \\
\theta_{AD, AC} + \theta_{AE, AC} < \pi, & \text{if } \vec c_3 \cdot \vec c_4 < 0.
\end{cases}
\]

\begin{wrapfigure}{l}{0.18\textwidth}
    \vspace{-1em}
    \centering
    \includegraphics[width=0.18\textwidth]{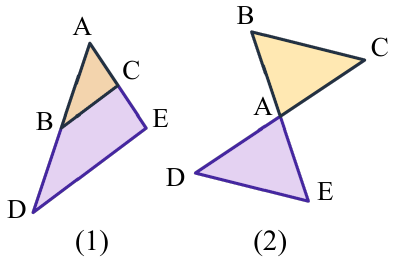}
    \vspace{-2em}
    \caption{Zero cross-product cases.}
    \label{fig:counterex2d}
    \vspace{-2em}
\end{wrapfigure}

To distinguish edge cases where the cross product is zero, as illustrated in Figure~\ref{fig:counterex2d} cases (1) and (2), we can utilize the sign of the dot product. Specifically, for cases where the cross product is zero, we can compare the sign of the dot product to determine the intersection.

For vectors \( \vec{u} \) and \( \vec{v} \), if \( |\vec{u} \times \vec{v}| = 0 \), we further compare the signs of the dot products to ensure accurate intersection checks.

The edge case handling can be described as follows:
$
\text{If } |\vec{u} \times \vec{v}| = 0, \text{ then compare } \text{sign}(\vec{u} \cdot \vec{w}) \text{ and } \text{sign}(\vec{v} \cdot \vec{w}) \text{ for a vector } \vec{w}.
$

\paragraph{Edge sharing case}
When two triangles share two vertices, \( A \) and \( B \), we propose the \textit{same-side test} to detect intersections. This test involves extending the shared edge \( AB \) into a line and determining the positions of the remaining vertices \( C \) and \( D \).
\begin{wrapfigure}{l}{0.18\textwidth}
    \centering
    \vspace{0.0em}
    \includegraphics[width=0.22\textwidth]{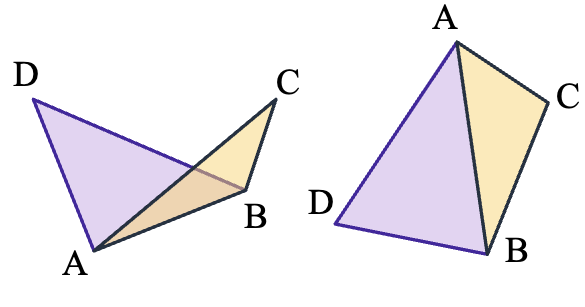}
    \vspace{-2.5em}
    \caption{Edge-sharing cases.}
    \label{fig:edge_vertex}
    \vspace{-5.0em}
\end{wrapfigure}
If \( C \) and \( D \) lie on opposite sides of the line \( AB \), it indicates a simple edge sharing. Otherwise, if they are on the same side, it indicates a self-intersection.



\vspace{2.0em}
\subsection{3D triangle intersection analysis}

Drawing on the capabilities of \citesupp{devillers2002faster}, which allows for the determination of the intersection line when triangles intersect robustly, we propose a method to manage these intersections based on the length of the intersection line. If two triangles share a vertex and the intersection line is very short, the intersection likely occurs at the shared vertex, and thus, should not be considered a self-intersection. Conversely, if the intersection line is long despite the vertices being shared, it suggests that the triangles, although sharing vertices, do intersect themselves. Since in 3D scenarios, triangles sharing an edge cannot occur intersection, we can simply skip the case.



\let\etalchar\relax

\bibliographystylesupp{eg-alpha-doi}

\end{document}